\documentclass[lettersize,journal]{IEEEtran}
\usepackage{amsmath,amsfonts}
\usepackage{algorithm}
\usepackage{array}
\usepackage[caption=false,font=normalsize,labelfont=sf,textfont=sf]{subfig}
\usepackage{textcomp}
\usepackage{stfloats}
\usepackage{url}
\usepackage{verbatim}
\usepackage{graphicx}
\usepackage{cite}
\usepackage{makecell}

\usepackage[numbers,sort&compress]{natbib}
\usepackage[noend]{algpseudocode}  
\usepackage{subcaption}

\begin{document}

\title{Tyche: A Hybrid Computation Framework of Illumination Pattern for Satellite Beam Hopping}

\author{Ziheng Yang, Kun Qiu, \IEEEmembership{Senior Member,~IEEE,}, Zhe Chen, \IEEEmembership{Member,~IEEE,},  Wenjun Zhu, \IEEEmembership{Member,~IEEE,}, and Yue Gao, \IEEEmembership{Fellow,~IEEE}

}
\markboth{IEEE Journal on Selected Areas in Communication}%
{Shell \MakeLowercase{\textit{et al.}}: A Sample Article Using IEEEtran.cls for IEEE Journals}


\maketitle

\begin{abstract}
High-Throughput Satellites (HTS) use beam hopping to handle non-uniform and time-varying ground traffic demand. A significant technical challenge in beam hopping is the computation of effective illumination patterns. Traditional algorithms, like the genetic algorithm, require over 300 seconds to compute a single illumination pattern for just 37 cells, whereas modern HTS typically covers over 300 cells, rendering current methods impractical for real-world applications. Advanced approaches, such as multi-agent deep reinforcement learning, face convergence issues when the number of cells exceeds 40. In this paper, we introduce Tyche, a hybrid computation framework designed to address this challenge. Tyche incorporates a Monte Carlo Tree Search Beam Hopping (MCTS-BH) algorithm for computing illumination patterns and employs sliding window and pruning techniques to significantly reduce computation time. Specifically, MCTS-BH can compute one illumination pattern for 37 cells in just 12 seconds. To ensure real-time computation, we use a Greedy Beam Hopping (G-BH) algorithm, which provides a provisional solution while MCTS-BH completes its computation in the background. Our evaluation results show that MCTS-BH can increase throughput by up to 98.76\%, demonstrating substantial improvements over existing solutions.
\end{abstract}

\begin{IEEEkeywords}
High-throughput Satellite, Beam Hopping, Monte Carlo Tree Search, Resource Management
\end{IEEEkeywords}

\section{Introduction}
\IEEEPARstart{W}{ith} the development of satellite technology, current Geostationary Earth Orbit (GEO) satellites are mostly High-Throughput Satellites (HTS) with multiple point beams~\cite{HTS}. Due to the non-uniform distribution of the ground population and the time-varying nature of the ground traffic demand, HTS using beam hopping can flexibly allocate beam resources to effectively match ground traffic demand. Beam hopping means adaptive activation of beams according to the actual traffic demand~\cite{BHADV,BHADV2}, utilizing beam resources in the form of time division multiplexing to serve ground cells. Several studies shows that in HTS systems, the use of beam hopping can reduce total payload power requirement, increase useable capacity and reduce unmet traffic demand~\cite{BH_INCREASE_Th2}.

One of the technical challenges in beam hopping is the computation of illumination patterns. Illumination patterns determine the different sets of beams that need to be activated in each time slot~\cite{NPref}. Multiple illumination patterns form a Beam Hopping Transmission Plan(BHTP). Upon receiving ground traffic demand collected by satellites, the Network Operation and Control Center (NOCC) must rapidly compute BHTP to enable satellites to promptly adjust beam scheduling strategies in response to changing traffic demands. The shorter the computation time for BHTP or individual illumination patterns, the faster satellites can update their scheduling strategies, thereby enhancing system throughput. Experiments show that appropriate beam scheduling strategies can improve system throughput by up to 98\%.

In the computation of illumination patterns, Co-Channel Interference (CCI)~\cite{CCI} and cell traffic demand are considered to achieve high throughput, and beam activation is usually represented by binary variables. So the computation of illumination patterns is typically modeled as a Mixed-Integer Nonlinear Programming (MINLP) problem~\cite{MINLP}, which is NP-hard and difficult to solve. Additionally, the number of cells covered by the current HTS can reach up to 300~\cite{Viasat,Konnect}, such as the JUPITER 3 satellite launched in 2022~\cite{JUPITER3}. The larger number of cells further increases the difficulty of solving this problem.

The most naive methods for illumination pattern computation are heuristic algorithms, which are typically based on greedy strategies~\cite{DyBhQifa1,DyBhQifa2,ZhanglEO,SNS3}. However, the current work does not provide theoretical upper and lower bounds for this type of algorithm, which makes it difficult to guarantee its performance. To ensure the performance of the algorithm, some studies use metaheuristic search methods such as the Genetic Algorithm (GA)~\cite{DyBhGA,GA1}. Nevertheless, such methods often have excessively high computational complexity. In our experiments, compared with heuristic algorithms, using GA in 37 cells can increase throughput by up to 26.19\%, but it requires more than 330s to compute an illumination pattern, which is unacceptable. 

The latest method is based on Deep Reinforcement Learning (DRL)~\cite{DRL1,DRL2,DRL3,MADRL,DRLGA}. Multi-Agent Deep Reinforcement Learning Beam Hopping (MADRL-BH) can achieve the same throughput performance as GA and its computation time meets the requirements for real-time computation~\cite{MADRL}. However, the performance of this method is validated only in scenarios of 19 cells. Through experiments, we verify that when the number of cells exceeds 40, due to the increase in the number of cells and beams, the model cannot learn effective strategies and converge in complex environments. So this type of method is also unsuitable for scenarios with a large number of cells.

To overcome those limitations and compute illumination patterns in large-scale cell scenarios, we propose Tyche, a hybrid computation framework of illumination pattern. Tyche uses a Monte Carlo Tree Search (MCTS) based algorithm~\cite{MCTS}, MCTS-BH, to compute illumination patterns. MCTS-BH can maximize throughput while reducing the computation time. Compared with other algorithms, the throughput of MCTS-BH can increase by up to 98.76\%. To reduce the computational complexity, we apply the sliding window algorithm and pruning algorithm to MCTS-BH, which reduces the computation time by up to 81.41\%. Although MCTS-BH can achieve high throughput, its computation time is still too long to meet the requirements of real-time computation. Therefore, We use a Greedy Beam Hopping (G-BH) algorithm to provide a provisional solution while MCTS-BH completes its computation in the background.

Briefly speaking, this paper makes the following contribution:
\begin{itemize}
\item We analyze the reasons why current research cannot be used in large-scale cells, and conduct experiments to verify that MADRL fails to converge in large-scale cells, as well as the problem of high computational complexity of traditional algorithms in large-scale cells.
\item We propose Tyche, a hybrid computation framework of illumination pattern. This framework can be used for the real-time illumination pattern computation in large-scale cells.
\item We design MCTS-BH to compute illumination patterns in Tyche and apply the sliding window algorithm and pruning algorithm to MCTS-BH to reduce computational complexity. In this way, the computation time is reduced by up to 81.41\%  while ensuring the same throughput. To satisfy the requirement of real-time computation, we use G-BH to provide a provisional solution while MCTS-BH completes its computation in the background.
\item To verify the throughput and computation time advantages of MCTS-BH, we conduct experiments for comparison with various benchmark algorithms.

\end{itemize}

The rest of the paper is organized as follows: Section II
introduces the background of the beam hopping. Section III demonstrates the overview design of Tyche. Section IV describes the MCTS-BH. Section V explains the optimization of MCTS-BH. Section VI evaluates G-BH and MCTS-BH and analyzes the evaluation results. Finally, Section VII provides the conclusion.

\section{BACKGROUND}
\subsection{Beam Hopping}
\begin{figure}
  \centering
  \includegraphics[width= \linewidth]{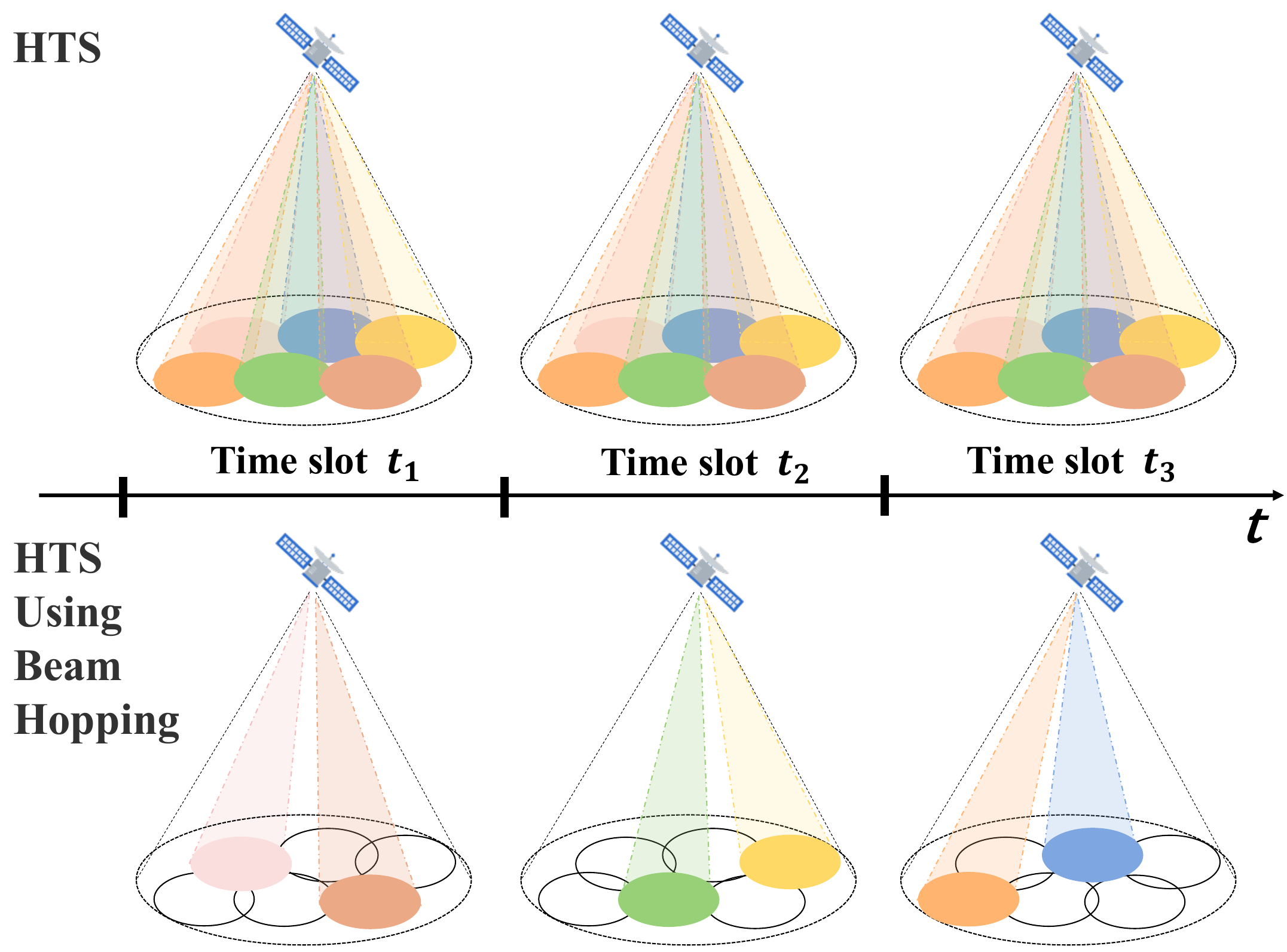}
  \caption{Comparison of HTS and HTS using beam hopping service modes. HTS using beam hopping can more flexibly schedule beam resources and select different service cells in different time slots according to demand.}
  \label{fig:beam compare}

\end{figure}

The scenario studied in this paper is a Ka-band GEO HTS system with multi-spot beams. In such systems, HTS typically achieves coverage by utilizing multiple spot beams, with each beam covering a single cell. Frequency reuse is implemented across cells to enhance spectral efficiency. However, due to the non-uniform distribution of the population on the ground, the traffic demand between cells has a non-uniform distribution. This causes an imbalance in supply and demand between cells. HTS using beam hopping selects only a few cells within the coverage range for service in a certain time slot. To fully utilize spectrum resources, each beam uses the same frequency and avoids CCI through spatial isolation. The difference in beam illumination between HTS using beam hopping and ordinary HTS as shown in Fig.~\ref{fig:beam compare}.

\subsection{Illumination Pattern}
\begin{figure}
  \centering
  \includegraphics[width=\linewidth]{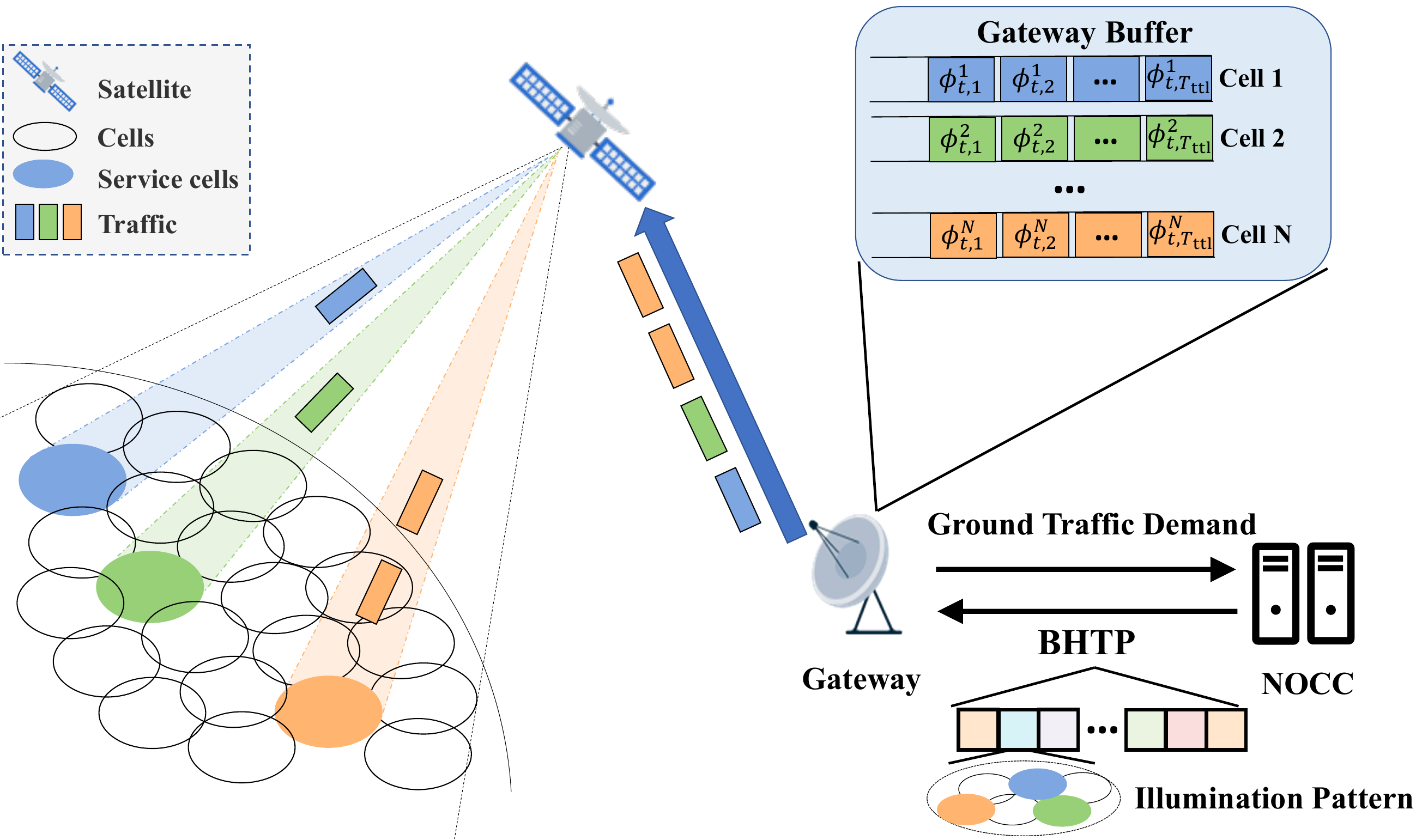}
  \caption{An example of a beam hopping system, where the NOCC calculates the BHTP and sends it to the satellite. The satellite manages the beams according to the BHTP.}
  \label{fig:system_arch}
 
\end{figure}

An illumination pattern refers to the set of service cells during a time slot. A BHTP consists of multiple such illumination patterns, each corresponding to a different time slot. Upon receiving ground traffic demand collected by the satellite, the NOCC computes the BHTP tailored to the current demand distribution and transmits it back to the satellite. The satellite then schedules its beams according to the BHTP, thereby enabling demand-aware beam allocation.The overall architecture of the beam hopping satellite system is shown in Fig. \ref{fig:system_arch}.

The computation of an illumination pattern must jointly consider both demand-driven beam allocation and inter-beam interference~\cite{11148638}. Balancing these two factors effectively is critical to the success of the algorithm. Notably, in the absence of an updated BHTP, the satellite continues scheduling beams based on the previously received BHTP. If the computation of the new BHTP is excessively time-consuming, the satellite cannot respond promptly to changes in traffic demand, resulting in suboptimal beam allocation.
 
In our experiments, the system throughput gap between appropriate and inappropriate BHTPs can be as high as 98.76\%. This shows how quickly calculating illumination patterns and generating BHTP greatly boosts the system's overall performance.

\subsection{Time Slot}
In this paper, a time slot is defined as the duration during which a satellite beam serves a fixed set of cells, with beam switching occurring between different time slots. According to the DVB-S2X~\cite{DVB-S2X}, superframes are used as the data transmission frame format (a frame structure specifically designed for beam-hopping scenarios in DVB-S2X, with a fixed frame length of 612,540 symbols), and several superframes are transmitted within each time slot. Taking typical DVB-S2X parameters as an example (roll-off factor of 0.25 and a bandwidth of 500 MHz), the symbol rate \( R_s \) is 400 MSym/s, resulting in a superframe duration of approximately 1.5 ms.

\begin{figure}
  \centering
  \includegraphics[width=0.7\linewidth]{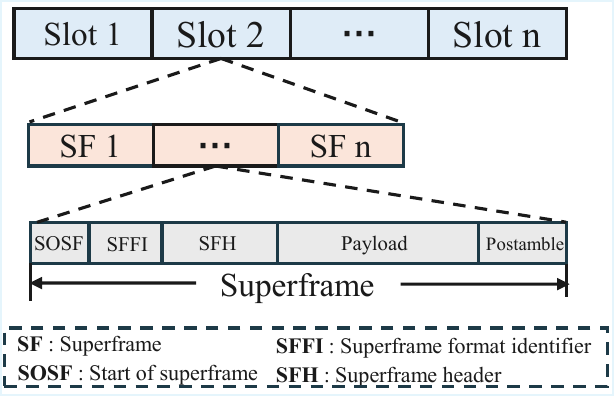}
  \caption{Time slot structure illustrating multiple superframes transmitted within a single timeslot.}
  \label{fig:Time slot}
 
\end{figure}

Due to the significant path loss in GEO satellite systems, excessively frequent beam switching would hinder terminal synchronization. Therefore, in this paper, each time slot is set to 100 ms, with one synchronization signal transmitted per slot. This configuration not only meets the requirement in GEO satellite scenarios that beam switching frequency should not be too high (each slot can transmit approximately 66 superframes), but also reserves about 1 ms as a guard interval to maintain frame boundaries and absorb clock drift and command jitter. At the same time, this setting complies with 3GPP Rel-19 specifications (less than 160 ms). In 3GPP Rel-19, it is proposed that, in beam-hopping scenarios, the Synchronization Signal Block (SSB) period can be configured to a maximum of 160 ms~\cite{3GPP2023}. In addition, a 100 ms time slot length is also a commonly used parameter configuration in current academic literature~\cite{DRL1}. The concept of the time slot is illustrated in Fig.~\ref{fig:Time slot}.

It should be emphasized that the time slot length in this paper is not a fixed value and can be adjusted according to specific system requirements. It is not a key determinant of the algorithm performance. The focus of this study is on algorithm design to maximize system throughput under fixed time-slot conditions. In practical deployments, the time slot length can be dynamically adjusted according to system requirements to accommodate 3GPP, DVB-S2X, or other standards.

\subsection{Previous Methods and Issues}

The simplest computation method for illumination patterns is the heuristic algorithm based on greedy strategy~\cite{DyBhQifa1,DyBhQifa2}. Although these methods have fast computation speeds, the existing research does not prove the theoretical upper and lower bounds of this algorithm, so it cannot guarantee that the algorithm obtains the global optimal solution. Through experiments, we verify that the throughput obtained by heuristic algorithms is not the highest.

Some studies also use metaheuristic search algorithms, such as GA~\cite{DyBhGA}. This type of algorithm requires excessively long computation time. In our experiments, the computation time for GA exceeds 300 seconds in 37 cells. Due to the excessively long computation time, it is also unsuitable for scenarios with a large number of cells. 

In DRL-based methods, some research highlights the issue of explosion in the DRL action space~\cite{DRL1}, which restricts the applicability of DRL in scenarios with a large number of cells. In MADRL, each beam is abstracted as an agent, thereby reducing the dimension of action space~\cite{MADRL}. This method can address the issue of action space explosion in traditional DRL-based methods. After pre-training, MADRL can achieve the same throughput performance as GA and the computation time is reduced by 6299x in scenarios with 19 cells. However, the MADRL performance is only validated in 19 cells. We verify that the MADRL model fails to converge when the number of cells exceeds 40. This is mainly because in complex environments, it is challenging for agents to learn effective strategies and achieve convergence. So DRL-based methods are also unsuitable for scenarios involving a large number of cells.

To address the problem that the above research is not applicable to large-scale cells, we propose Tyche to solve the illumination pattern computation problem in large-scale cells. We design MCTS-BH in Tyche and reduce the computation time of MCTS-BH through the sliding window algorithm and pruning algorithm, but it still does not meet the requirements for real-time computation. Therefore, we use G-BH to provide a provisional solution during MCTS-BH computation. 

\begin{figure*}
  \centering
  \includegraphics[width=0.9 \linewidth]{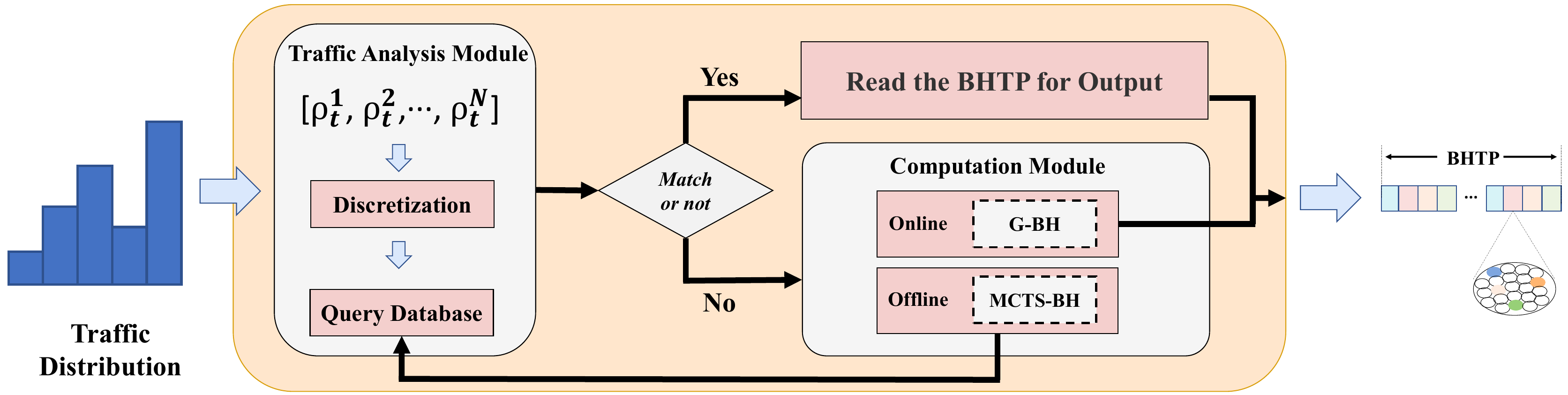}
  \caption{The overview workflow of Tyche, Tyche consists of a traffic analysis module and a computation module. The traffic analysis module is used to query the local database for the precomputed BHTP. The computation module computes the BHTP, utilizing both MCTS-BH and G-BH simultaneously. The result of G-BH is output directly, while the result of MCTS-BH is stored in the database.}
  \label{fig:jeoy architecture}
 
\end{figure*}

\section{DESIGN OVERVIEW}

To enable real-time illumination pattern computation, we design Tyche, which utilizes the G-BH algorithm for the online computation and the MCTS-BH algorithm for the offline computation. As shown in the  Fig.~\ref{fig:jeoy architecture}, Tyche includes a traffic analysis module and a computation module. When computing a BHTP, Tyche first uses the traffic analysis module to analyze the current traffic distribution. It then queries the local database to determine if there is a BHTP corresponding to this traffic distribution. If a matching BHTP is found, Tyche outputs it directly. Otherwise, Tyches uses the online algorithm G-BH to compute multiple illumination patterns to form the BHTP output. Simultaneously, Tyche uses MCTS-BH as the offline computation algorithm to compute illumination patterns to form the BHTP and stores it in the local database, so that the BHTP can be directly returned the next time the same traffic distribution is requested.

\subsection{Traffic Analysis Module}

\begin{figure}
  \centering
  \includegraphics[width= 0.7\linewidth]{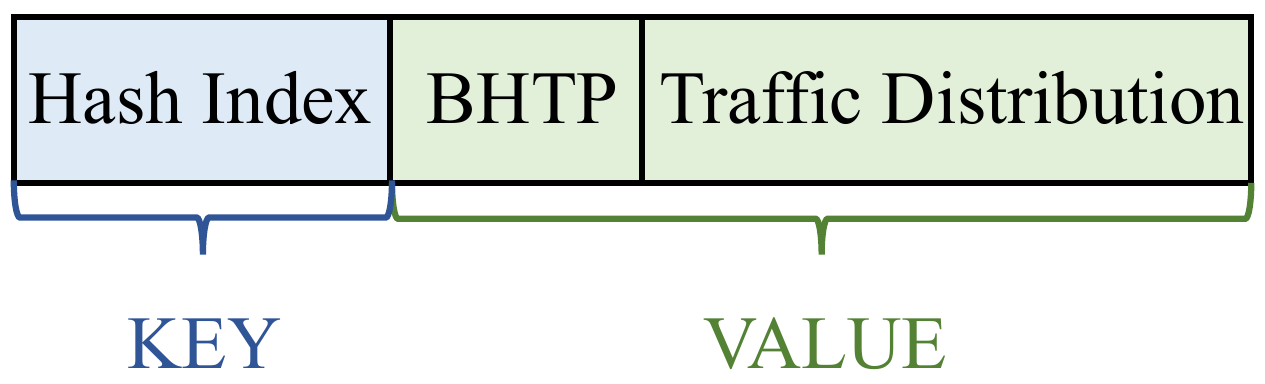}
  \caption{The database storage format for the traffic analysis module}
  \label{fig:data_structure}
\end{figure}

This module is used to determine whether the current database stores a BHTP corresponding to the ground traffic distribution. In the specific implementation, a hash-based database such as Redis can be used to efficiently store ground traffic distributions and the corresponding BHTP. For example, a set of traffic distribution vectors can be processed using a hash function (such as SHA-256) to generate a unique key, which enables quick lookup of database records. The value part of the database includes the BHTP and the original traffic distribution vector to support verification in case of hash collisions. The design of the database storage format is illustrated in the Fig.~\ref{fig:data_structure}. Because of the in-memory operations and hash table efficiency of Redis, the query time complexity is O(1). Benchmark tests indicate that on a standard computer, the latency of a single query operation is typically within 1 ms \cite{RedisBenchmarking}, fully meeting real-time requirements.

Although ground traffic distribution is fundamentally a continuous variable, its discretization becomes feasible and advantageous due to the illumination pattern calculation relying on the relative traffic magnitudes across different cells. This approach enables efficient storage of traffic distributions, reducing memory usage while enhancing the hit rate. Specifically, we define the maximum beam capacity as \( C_{\text{max}} \) and introduce a discretization factor \( \beta \) to regulate the granularity of the discretization process. The discretized traffic values are represented by an evenly spaced discretization grid, defined as follows:

\begin{equation}
\left\{ k \cdot \Delta C \;\middle|\; \Delta C = \frac{C_{\text{max}}}{\beta},\; k = 0, 1, 2, \ldots, \beta \right\}
\end{equation}

Here, the discretization factor \( \beta \) serves as a key parameter to balance memory usage and performance: a higher \( \beta \) results in finer discretization, improving matching accuracy and enabling more precise illumination pattern calculations, which in turn enhances throughput. However, this also leads to increased memory consumption. In the traffic analysis module, Tyche maps the traffic distribution to a discrete value within this grid, generating a discretized representation of the traffic distribution. This method significantly reduces the number BHTPs that need to be stored, thereby optimizing memory efficiency and improving the query hit rate.

\subsection{Computation Module}
This module is mainly used to compute illumination patterns. When there is no corresponding BHTP in the database, Tyche uses the computation module to compute it. This module contains two algorithms:an online computing algorithm and an offline computing algorithm. Tyche will use the online computing algorithm G-BH to quickly compute illumination patterns for BHTP output. Simultaneously, Tyche uses the offline algorithm MCTS-BH to compute illumination patterns forming the BHTP and stores it in the database. G-BH can compute an illumination pattern within a short time, but it does not achieve the highest throughput. MCTS requires a longer computation time but can achieve a higher throughput. Therefore, the results computed by MCTS-BH are stored in the database, which ensures that the database contains illumination patterns that can achieve a higher throughput. The hybrid computation design ensures that Tyche maintains short computation times while maximizing the throughput as much as possible.

\section{ALGORITHM DESIGN}
In this section, we introduce the system model and the details of MCTS-BH.

\subsection{Problem Formulation}
This paper explores a forward link scenario for a GEO satellite employing beam hopping, with the system schematic illustrated in Fig.~\ref{fig:system_arch}. The satellite generates \( K \) beams through time-division multiplexing to serve \( N \) ground cells. Due to uneven population distribution and time-varying user traffic, the traffic demand of each cell varies. To store this traffic, the gateway is equipped with \( N \) queues, each corresponding to a specific cell, to hold the downlink traffic data for that cell. At time slot \( t \), the total traffic stored in queue \( n \) is denoted as \( d_t^n \). Within queue \( n \), the number of data packets that have been waiting for \( l \) time slots is represented by \( \phi_{t,l}^{n} \). It is assumed that each data packet can only be stored in the queue for \( T_{ttl} \) time slots, where \( T_{ttl} \) is defined as the time-to-live (TTL). Thus, we have \( d_t^n = \sum_{l=1}^{T_{ttl}} \phi_{t,l}^{n} \). The arrival traffic rate for queue \( n \) at time slot \( t \) is denoted as \( \rho_n^t \), and its magnitude depends on the traffic demand of cell \( n \). In this scenario, the illumination pattern in time slot $t$ can be expressed as
\begin{equation}
    X_t = \{ (x_t^1, x_t^2, ..., x_t^n, ..., x_t^N) \mid x_t^n = 0, 1 \text{ and } \sum_{i=1}^{N} x_i^t = K \},
\end{equation}
where $x_t^n = 1$ indicates that cell $n$ is illuminated in time slot $t$.

In each cell, a single user represents the total traffic demand of all users within that cell (since multiple access techniques within the cell are not the focus of this work). The channel coefficient between beam \( k \) and the user in cell \( n \) (i.e., user \( n \)) can be expressed as
\begin{equation}
\begin{aligned}
h_{k,n} = \frac{\sqrt{G_t(\theta_{k,n}) G_r^n}}{4 \pi \frac{d_{k,n}}{\lambda}},
\end{aligned}
\end{equation}
where \( G_t(\theta_{k,n}) \) is the transmit antenna gain from beam \( k \) to user \( n \), \( G_r^n \) is the receive antenna gain of user \( n \), \( d_{k,n} \) represents the spatial distance between the satellite and user \( n \), and \( \lambda \) is the wavelength of the propagated signal.

Therefore, the signal-to-interference-plus-noise ratio (SINR) for user \( n \) in time slot \( t \) can be expressed as
\begin{equation}
\begin{aligned}
\mathrm{SINR}_{n}^t = \frac{P_k |h_{k,n}|^2}{k_B T_{\mathrm{rx}} B_k + \sum_{l \in S} P_l |h_{l,n}|^2} 
\label{eq:sinr}
\end{aligned}
\end{equation}
where \( P_k |h_{k,n}|^2 \) represents the received signal power, with \( P_k \) being the transmit power of beam \( k \). The term \( k_B T_{\mathrm{rx}} B_k \) denotes the thermal noise power of the receiver, where \( k_B \) is the Boltzmann constant, \( T_{\mathrm{rx}} \) is the receiver noise temperature, and \( B_k \) is the signal bandwidth. The summation term \( \sum_{l \in S} P_l |h_{l,n}|^2 \) represents the cumulative interference power from co-frequency beams, where \( S \) is the set of beams sharing the same frequency as beam \( k \).
The offered traffic (available throughput) of cell $n$ can be computed as
\begin{equation}
    C_t^n = x_t^nB_k\log_2 (1 + \text{SINR}_n^t),\
\label{equ:shannon}
\end{equation}
where $B_k$ represents the bandwidth of beam $k$. Thus, the data throughput for cell $n$ in time slot $t$ is
\begin{equation}
    \omega_t^n = \min \{C_t^nT_{slot}, d_t^n\Lambda\},
\label{equ:Th}
\end{equation}
where $T_{slot}$ represents the duration of one time slot. $\Lambda$ represents the number of bits in each data packet.Therefore, the change in the data volume in queue \( n \) can be expressed as:
\begin{equation}
    d_t^n = d_{t-1}^n - \omega_{t-1}^n + \rho_n^t
    \label{eq:queue change}
\end{equation}

Our goal is to effectively utilize satellite resources and maximize the throughput of the satellite. Therefore, we establish an optimization model with the goal of maximizing throughput, which can be expressed as:
\begin{equation}
\begin{aligned}
    \text{opt.} \max \quad &\sum_{t=1}^{T} \sum_{n=1}^{N} \omega_t^n\\
    \text{s.t.} \quad
    &C1: \sum_{i=1}^{K} P_{b}^{k} < P_{\text{tot}}, \quad \forall k \in K \\
    &C2: P_{b}^{k} < P_{\text{max}}, \quad \forall k \in K \\
    &C3: \sum_{i=1}^{N} x_{t}^{i} = K \quad \text{and} \quad x_{t}^{i} = 0 \text{ or } 1
\label{con:opt}
\end{aligned}
\end{equation}

In the above optimization problem, \( T \) is the number of time slots. In this study, we focus only on the throughput performance within T time slots. C1 requires the sum of all beam power is less than the total satellite power \( P_{\text{tot}} \), and C2 indicates the constraint for the maximum power of each beam \( P_{\text{max}} \). 
Because the power allocation is not discussed in this paper, the transmission power of each beam is evenly distributed. So the power of each beam is set to \( P_k^b = P_{\text{tot}}/K < P_{\text{max}} \). Thus, the constraints C1 and C2 can be satisfied. C3 indicates that the number of beams activated simultaneously should be equal to \( K \). Due to the influence of Equ. (\ref{equ:Th}) and Equ. (\ref{eq:sinr}), the problem manifests as a non-convex, nonlinear integer programming problem, which is proven to be NP-hard~\cite{MADRL}. In the following section, we will introduce MCTS-BH to solve this problem.

\begin{table}
\centering
\caption{Notation and Description}
\begin{tabular}{l|c}
\hline
Notation & Description \\
\hline
$N$ & Number of cells \\
$K$ & Number of beams \\
$P_b$ & Beam power \\
$P_{tot}$ & Total satellite power \\
$P_{max}$ & The maximum power of each beam \\
$T_{ttl}$ & Time to live of data packet in queue \\
$T_{slot}$ & The duration of one time slot \\
$T$ & The number of time slots \\
$d_t^n$ & The number of packets in queue \(n\) in time slot  \(t\) \\
$\phi_{t,l}^{n}$ & The number of packets that have waited \(l\) time slots \\
 & in queue \(n\) in time slot $t$ \\
\( \rho_n^t \) & The arrival traffic rate for queue \( n \) at time slot \( t \)  \\
$\Lambda$ & The number of bits in each data packet \\ 
$x_t^n$ & Whether the cell $n$ is illuminated by a beam in time slot $t$ \\
$C_t^n$ & The channel capacity of cell $n$ in time slot $t$ \\
$R_s$ & The symbol rate \\
$B_k$ & The bandwidth of beam $k$ \\
$\omega_t^n$ & The throughput of cell $n$  in time slot $t$ \\
$\mu_i$ & The selection value of node $i$ in the pruning algorithm \\
$D_{i,j}$ & The distance between cell $i$ and cell $j$ \\

\hline
\end{tabular}
\end{table}

\subsection{MCTS-BH Algorithm}
MCTS-BH applies the traditional MCTS algorithm to the beam hopping problem. When using MCTS, it is necessary to define the tree node, action, and score according to the problem. Below, we will explain these definitions and the process of MCTS-BH.

\subsubsection{Tree Node}
The computation of illumination patterns is selecting service cells in a particular time slot. So a tree node represents a selection of service cells. If the cell selection represented by a node has already selected \(K\) cells where \(K\) is the number of beams, the node is considered a terminal node. 
\subsubsection{Action}
Actions are defined as the selection of an unselected cell to be added to the set of selected cells. For a node, possible actions are unselected cells.
\subsubsection{Score}
In MCTS-BH, the simulation strategy is to randomly select unchosen cells until the node reaches the terminal state, which is inspired by the idea of a random walk in Go~\cite{alphaGo}. So the score in the simulation process is obtained by calculating the throughput represented by the terminal node. The score can be denoted as
\begin{equation}
        S^{'} =  \frac{\sum_{n=1}^{N} \omega_t^n } {\omega_{max}},  
\end{equation}
where \(\omega_{max} \) is the value used for normalization. The score of a node is the sum of the scores obtained from the simulation of that node and its child nodes. For node \( i\), its score can be defined as

\begin{equation}
        S_{i} =  \sum_{m \in I} S_{m}^{'}, \  
\end{equation}
where \( I\) is the set of nodes \( i\) and its child nodes. \(  S_{m}^{'}\) represents the score obtained from the simulation of node \( m\).

\begin{figure}
  \centering
  \includegraphics[width= \linewidth]{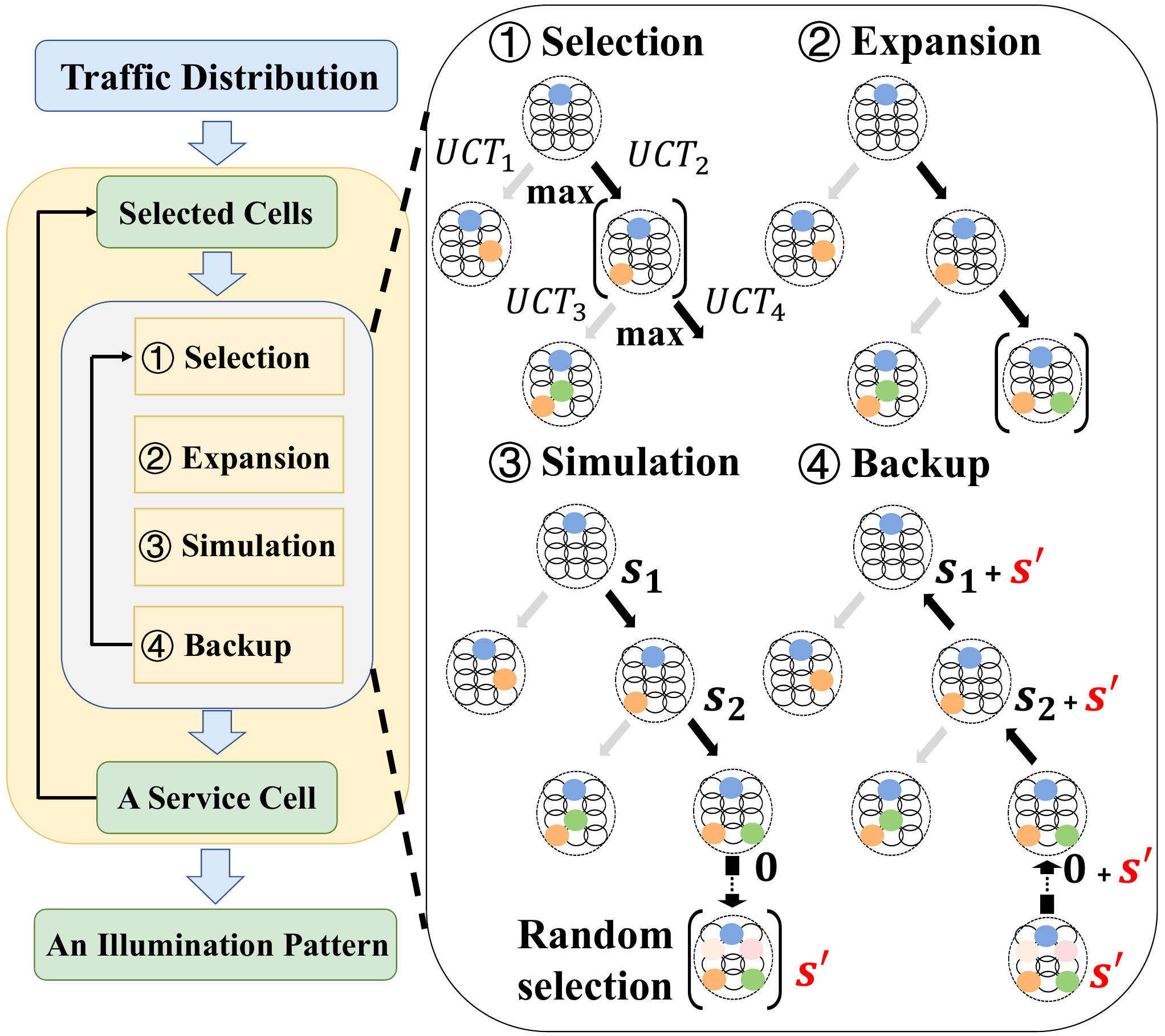}
  \caption{ The flowchart of MCTS-BH. In MCTS-BH, K MCTS iterations are performed, with each MCTS selecting one cell. It is worth noting that in the simulation and backup steps, the black text next to the tree node represents the original score of that node, and the red text represents the score obtained during the simulation process.}
  \label{fig:MCTS2}
 
\end{figure}

\subsubsection{The Process of MCTS-BH}
MCTS-BH sequentially executes the MCTS algorithm \(K\) times to determine the service cells for \(K\) beams. An unselected cell is chosen for each MCTS. After adding this cell to the set of selected cells, the next MCTS iteration is performed. After \(K\) MCTS iterations, we obtain \(K\) selected cells, which are the cells to be served in that time slot. The flowchart of the algorithm is shown in Fig.~\ref{fig:MCTS2}. 

Each MCTS has four steps: selection, expansion, simulation, and backup. Each MCTS loops through these four steps until it reaches its maximum number of iterations. The MCTS process is the same as that of the traditional MCTS, therefore, we do not elaborate on it.

\section{MCTS-BH Optimization}

To reduce the computational complexity of MCTS-BH, we optimize it from two aspects. On the one hand, we propose a scoring algorithm based on sliding window in the simulation process of MCTS, reducing the computational complexity of the scoring part from \( O(n^{2})\) to \( O(n)\). On the other hand, we designe a pruning algorithm to be applied in the expansion process of MCTS, accelerating the convergence of the algorithm. This section provides a detailed introduction to them.

\subsection{Scoring Algorithm Based on the Sliding Window Algorithm}
\begin{figure}[ht]
  \centering
  \includegraphics[width= 0.8 \linewidth]{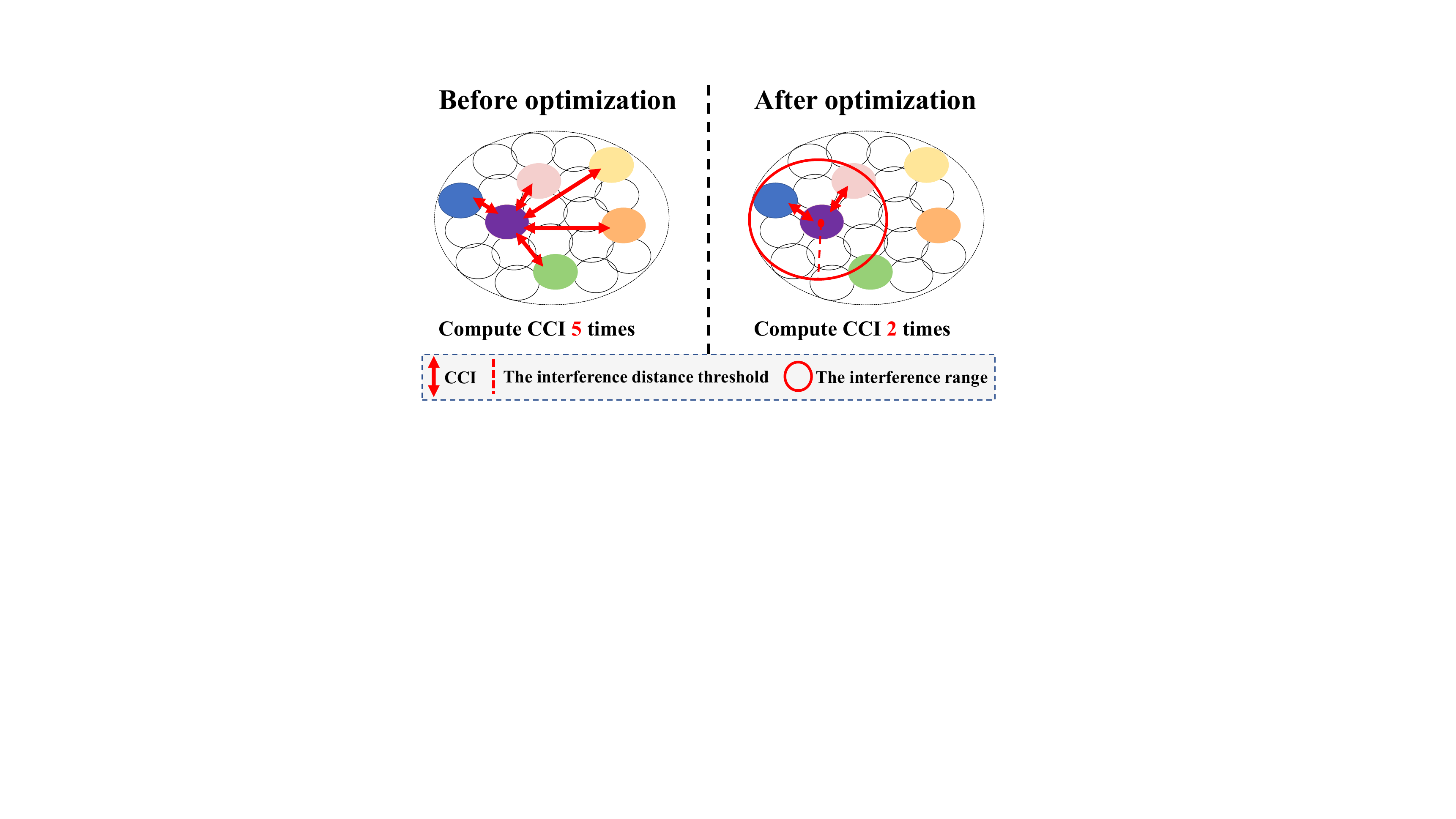}
  \caption{ Comparison of the number of CCI computations before and after optimization. Before optimization, computing the CCI for a cell requires computing the interference from all other service cells. After optimization, only the interference within the interference range needs to be computed. This approach reduces the number of CCI computations by ignoring smaller interference.}
  \label{fig:CCI}
 
\end{figure}
In the simulation process of MCTS, we need to score the illumination pattern represented by a terminal node. Because the scoring is performed at each simulation step, its speed greatly affects the speed of MCTS-BH. As mentioned earlier, the score is the throughput computed based on the illumination pattern. If the throughput is computed using the traditional method by \ref{equ:Th}, that is, when computing the SINR of a cell, it is necessary to compute the interference of all other service cells to that cell, which is why the computational complexity of the scoring algorithm is \( O(n^{2})\). In the scoring algorithm based on the sliding window algorithm, we obtain lower computational complexity by ignoring some small interference. When we choose to compute the SINR of a cell, we compute only the interference power within a fixed range of that cell. We define this range as the interference range, which is a circle with the cell as the center and the interference distance threshold \( D_{s}\) as the radius. The value of \( D_{s}\) is directly proportional to the accuracy and computational complexity of the algorithm. A specific example is shown in Fig.~\ref{fig:CCI}. But when counting interference cells within the interference range, if computing the distance between all other service cells and that cell each time, the computational complexity is still \( O(n^{2})\). Therefore, we use the sliding window algorithm to count the interference cells for each cell, and the computational complexity is \( O(n)\). 

\begin{figure*}
  \centering
  \includegraphics[width= 0.8\linewidth]{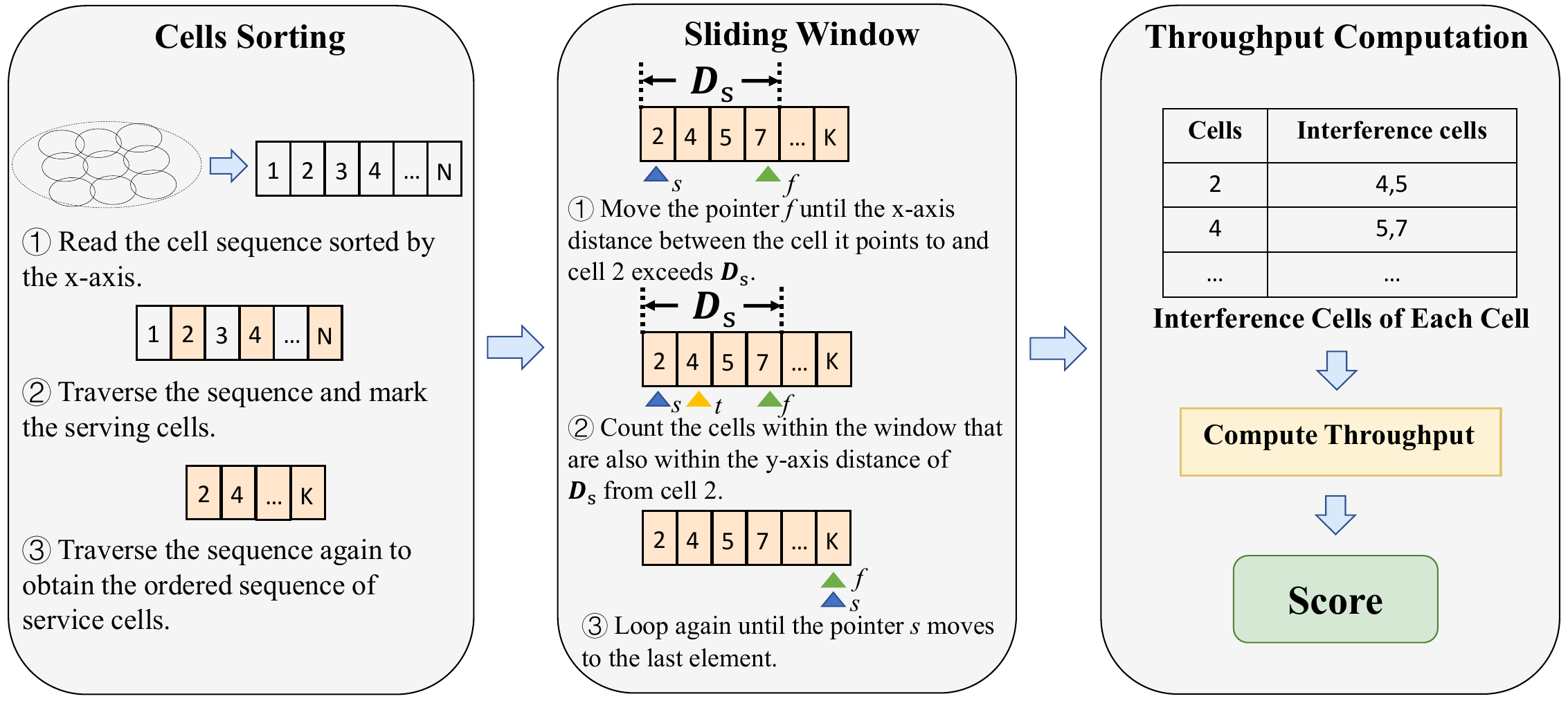}
  \caption{The overview workflow of the scoring algorithm based on the the sliding window algorithm. Firstly, all service cells of an illumination pattern are sorted according to their location information. Secondly, the sliding window algorithm is used to count interference cells of each service cell. Finally, the score is outputted according to interference cells of each service cell.}
  \label{fig:sliding window}
 
\end{figure*}

So the scoring algorithm based on the sliding window algorithm can be divided into three steps: sorting based on cell coordinates, sliding window algorithm for counting interference cells and computing the throughput of each cell, as shown in Fig.~\ref{fig:sliding window}. In the following, we introduce the details of each step.

\subsubsection{Sorting Based on Cell Coordinates}
To reduce the computational complexity of sorting, we leverage the fact that HTS is relatively stationary with respect to Earth. We pre-store an ordered sequence of cells within the HTS coverage area, with the sequence sorted by their x-axis coordinates. Here, we use the Mercator projection coordinates. This sorted cell sequence can be reused. Below, we introduce the specific process of sorting. First, traverse all cells in the illumination pattern and mark the service cells in the pre-sorted ordered cell sequence. Then, traverse the ordered cell sequence again, selecting the marked cells in order. Through these two traversals, we can obtain an ordered sequence of service cells.

\begin{algorithm}[t]
    \caption{Interference cells statistics based on the sliding window algorithm}
    \begin{algorithmic}[1]
        \Require ordered\ cells\ list \(\lambda\), interference distance threshold  \(D_{s}\)
         \Ensure a dictionary used to represent the interference cells for each cell $\beta$.
        
        \State $s$ and $f$ represent the start and end of a window
        \State $L$ is the size of \(\lambda\)
        
        \For{cell $c$ in \(\lambda\)}
            \State Add $c$ to $\beta$ with an empty list
        \EndFor
        
        \State $f \leftarrow 0$
        \State $s \leftarrow 0$
        
        \While{$s < L$}
            \While{$f < L$}
                \State $k \leftarrow$ x-axis distance between \(\lambda[f]\)  and  \(\lambda[s]\) 
                \If{$k \leq D_{s} $}
                    \State $f \leftarrow f + 1$
                \Else
                    \State \textbf{break}
                \EndIf
            \EndWhile
            
            \State $t \leftarrow s + 1$
            \While{$t \leq f$}
                 \State $k \leftarrow$ y-axis distance between \(\lambda[t]\)  and  \(\lambda[s]\) 
                \If{$k \leq D_{s} $}
                    \State Add $\lambda[t]$ to $\beta[\lambda[s]]$
                    \State Add $\lambda[s]$ to $\beta[\lambda[t]]$
                \EndIf
                \State $t \leftarrow t + 1$
            \EndWhile
            
            \State $s \leftarrow s + 1$
            \If{$f < s$}
                \State $f \leftarrow s$
            \EndIf

        \EndWhile
        
        \State \textbf{return} \(\beta\)
    \end{algorithmic}
\label{alg:sliding window}
\end{algorithm}

\subsubsection{The sliding Window Algorithm for Counting Interference Cells}
After obtaining the ordered sequence of service cells, the next step is to count the interference cells within the interference range of each cell. For this purpose, we use the sliding window algorithm. The algorithm employs three pointers: a slow pointer represents the start of the window, a fast pointer represents the end of the window, and a temporary pointer is used to iterate through the elements within the window. In each iteration, we count the interference cells of cell \( C_s\), which is indicated by the slow pointer. The cells within the window are considered possible interference cells. Once the window range is fixed, we use a temporary pointer to sequentially check the distance along the y-axis between cells within the window and \( C_s\)  to determine whether the cell is an interference cell for \( C_s\). The algorithm is shown in Algorithm \ref{alg:sliding window}.

\subsubsection{Computing the Throughput of Each Cell}
After obtaining the interference cells for each cell, we can compute the throughput of each cell. Using this as a basis for scoring.

\subsection{Pruning Algorithm}

MCTS expands the search tree through the expansion process. Expanding the search tree can prevent the algorithm from becoming stuck in local optima. But it also causes the MCTS algorithm to perform many iterations on low-value nodes. To enable the MCTS to perform more iterations on high-value nodes, we propose a pruning algorithm. In this algorithm, we define the selection value of an unselected node as
\begin{equation}
        \mu_{i} = \frac{d_t^n}{d_{max}} +  \frac {\sum_{j \in N'} D_{i,j}} {D_{max} }, \
\end{equation}
where  $\mu_{i}$ represents the selection value for node $i$. \(N'\) represents the set of selected cells, and  \(D_{i,j}\) represents the distance between cell \(i\) and cell \(j\), \(D_{max}\) and \(d_{max}\) represent the value used for normalization. This definition indicates that during MCTS expansion, it prefers to select a cell with a larger data volume in the traffic queue and that is farther away from the selected cells.

\begin{figure}
  \centering
  \includegraphics[width= 0.7\linewidth]{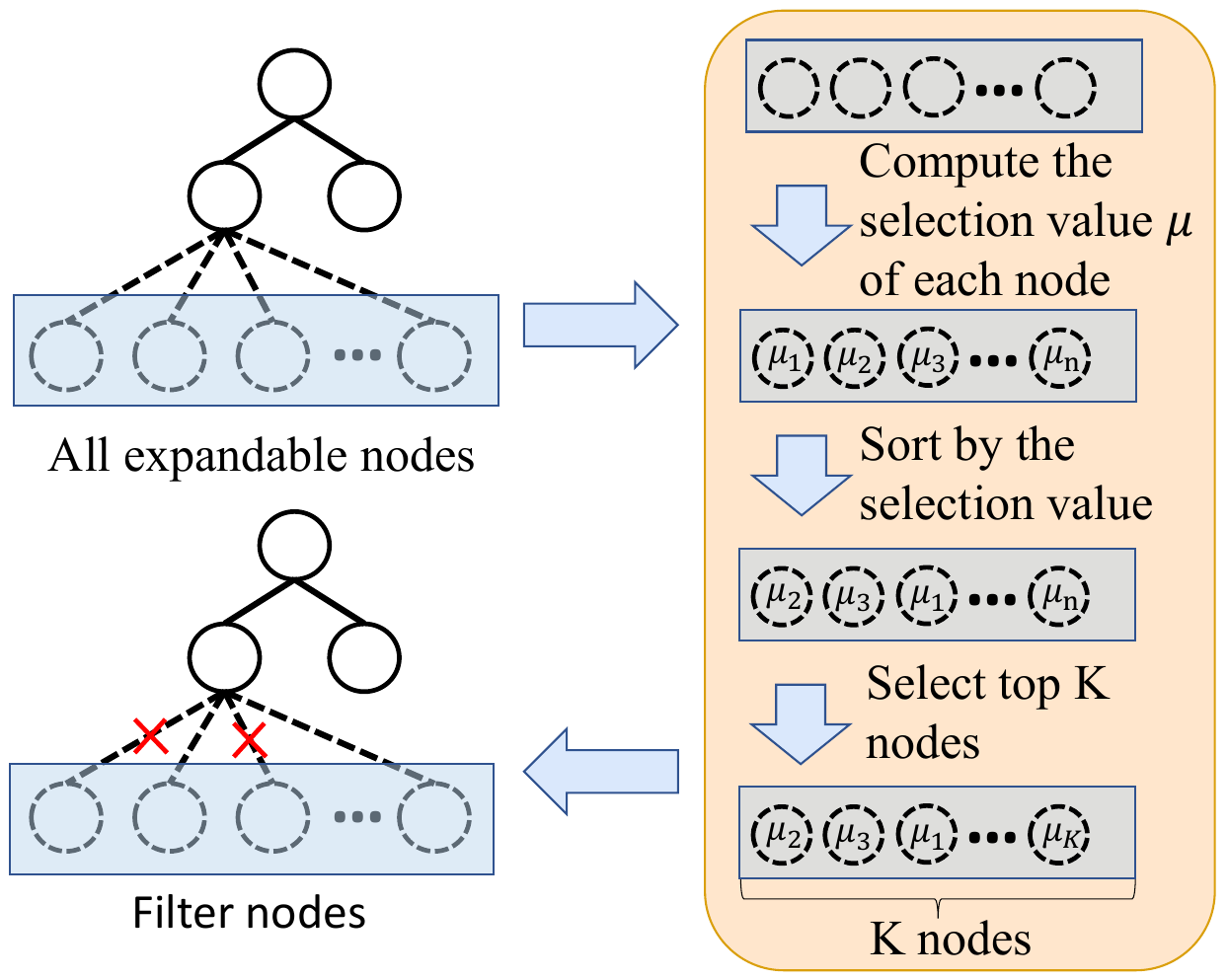}
  \caption{During expansion, MCTS-BH filters out points with lower selection value using a pruning strategy.}
  \label{fig:purning}
\end{figure}

During the expansion, the algorithm sorts the unselected cells according to their selection value, and then defines the top \(K\) cells with higher selection values as selectable actions. The process is shown in Fig.~\ref{fig:purning}. This is equivalent to using pruning strategies to filter out cells with a lower selection value, rather than using multiple simulations to filter out cells with a lower selection value, thereby reducing the computational complexity. The experimental results show that after applying pruning algorithm, MCTS-BH can achieve convergence with fewer iterations, and the throughput performance is basically the same as that of MCTS-BH without pruning.

\section{EVALUATION}
\subsection{Evaluation Parameter}
In this study,  we simulate a forward link communication scenario of a HTS using beam hopping in the Ka-band. We use the physical layer module in Plotinus~\cite{Plotinus} to simulate antenna and beam interference. We use the H3 library to generate cells~\cite{H3}. H3 provides the API to generate several hexagonal cells within a circle, with the number of cells determined by the input parameter. We use four parameters 3, 4, 5, and 6 to generate 37, 61, 91, and 127 cells, respectively. The number of beams is set to one-fourth of the number of cells, which is a typical ratio in previous studies~\cite{MADRL}. Therefore,  the number of cells covered by the satellite and the number of beams can be represented by ($37$, $9$), ($61$, $15$), ($91$, $22$) and ($127$, $31$). The duration of a time slot is set to 100ms, a value commonly adopted in related research on beam hopping~\cite{DRL1}. In practice, the time slot duration can be flexibly adjusted according to the requirements of different systems. The traffic demand of the cells is randomly generated, with a few cells randomly selected as hotspot cells to simulate non-uniform ground traffic demand. The antenna pattern is referenced from the 3GPP TR 38.811~\cite{3gpp}. Since the sidelobe level affects beam interference, thereby impacting system throughput, larger sidelobes require MCTS-BH to use a larger interference distance threshold \( D_{s}\) to ensure accurate scoring, resulting in longer computational times. Therefore, sidelobe amplitude is a critical factor determining algorithm performance. In this study, we adopt a generic antenna model, commonly used in academic research~\cite{antenna}, to evaluate the performance of MCTS-BH under conventional scenarios. The values of \( D_{s}\) are determined through multiple experiments. All parameters are listed in Table \ref{table:simulation parameters}.

\begin{table}
    \caption{EVALUATION PARAMETERS}
    \label{table:simulation parameters}
    \centering
    \begin{tabular}{|c|c|}
    \hline
    \rule[-6pt]{0mm}{18pt} Satellite altitude $H$ & 36000 km \\
    \hline
    \rule[-6pt]{0mm}{18pt} Frequency $f_c$ & 20 GHz \\
    \hline
    \rule[-6pt]{0mm}{18pt} \makecell{(Number of cells $N$, Number of beams $K$)} & \makecell{(37, 9), (61, 15),\\ (91, 22), (127, 31)} \\
    \hline
    \rule[-6pt]{0mm}{18pt} Beam power $P_b$ & 27 dBW \\
    \hline
    \rule[-6pt]{0mm}{18pt} 3 dB beamwidth $\theta_b$ & 1.5° \\
    \hline
    \rule[-6pt]{0mm}{18pt} Maximum transmit antenna gain $G_m$ & 40.3 dBi \\
    \hline
    \rule[-6pt]{0mm}{18pt} Terminal receiving antenna gain $G_{rx}$ & 31.6 dBi \\
    \hline
    \rule[-6pt]{0mm}{18pt} Time slot duration $T_{slot}$ & 100 ms \\
    \hline
    \rule[-6pt]{0mm}{18pt} Time to live of data packet in queue $T_{ttl}$ & 20 $T_{slot}$ \\
    \hline
    \rule[-6pt]{0mm}{18pt} The interference distance threshold \( D_{s}\) & 1 cells diameter \\
    \hline
    \end{tabular}
\end{table}

During the evaluations, we implement the following comparative algorithms:
\begin{enumerate}
    \item The Random Beam Hopping(R-BH): This method randomly selects $K$ cells as service cells in each time slot.
    \item The Periodic Beam Hopping(P-BH): This method periodically selects $K$ cells as service cells in each time slot, i.e. service cells are chosen in a round-robin manner.
    \item The Greedy Beam Hopping(G-BH): Select the K cells with the highest traffic demand as the service cells.
    \item MCTS-BH: Using the MCTS-BH algorithm for service cells selection.
    \item The Genetic Algorithm Beam Hopping(GA-BH): Using the genetic algorithm for service cells selection. The population is set to 500 and the number of generations is set to 50.
    \item MADRL-BH: According to~\cite{MADRL}, Using the MADRL algorithm for service cell selection.
\end{enumerate}

\subsection{The Issue of MADRL}

\begin{figure}
  \centering
  \includegraphics[width= 0.8\linewidth]{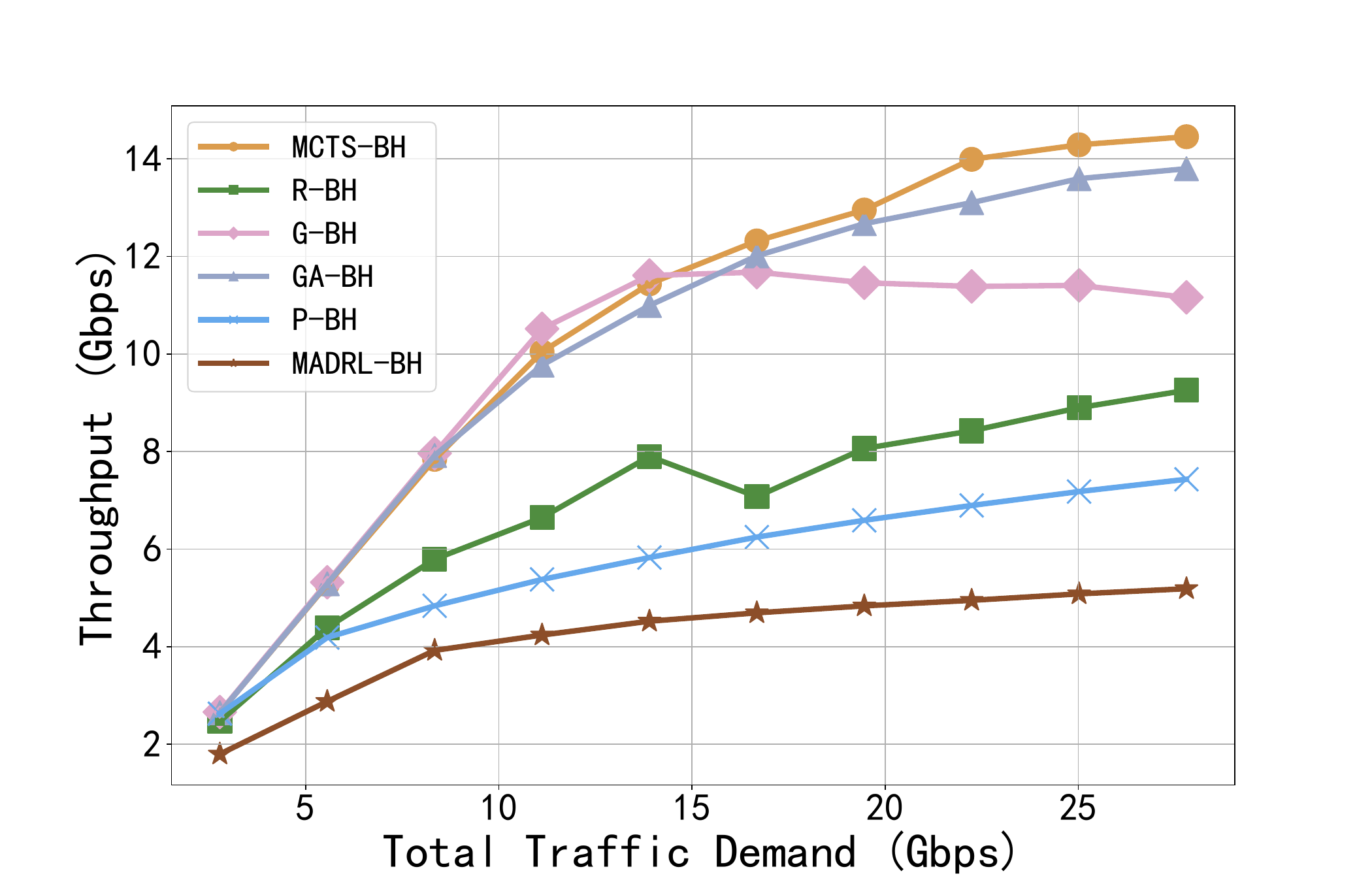}
  \caption{Comparison of throughput using different algorithms in 37 cells. MADRL-BH performs the worst in terms of throughput.}
  \label{fig:total_throughput_per_sim}
 
\end{figure}

We conduct experiments in a scenario of 37 cells and 9 beams, using various algorithms to compute illumination patterns for 30 time slots and compute the throughput in different traffic demand scenarios. The experimental parameters are set to be consistent with~\cite{MADRL}. The throughput performance of each algorithm under different traffic demand is shown in Fig.~\ref{fig:total_throughput_per_sim}. Because agents do not learn effective strategies, the performance of MADRL-BH is poor, and in some cases, it is even inferior to R-BH. This indicates that MADRL-BH is not suitable for large-scale cells scenarios, so we no longer use MADRL-BH as the baseline in a larger number of cells.

\subsection{Performance of Throughput}

\begin{figure*}
    \centering
    \subfloat[61 cells]{\includegraphics[width=0.31\linewidth]{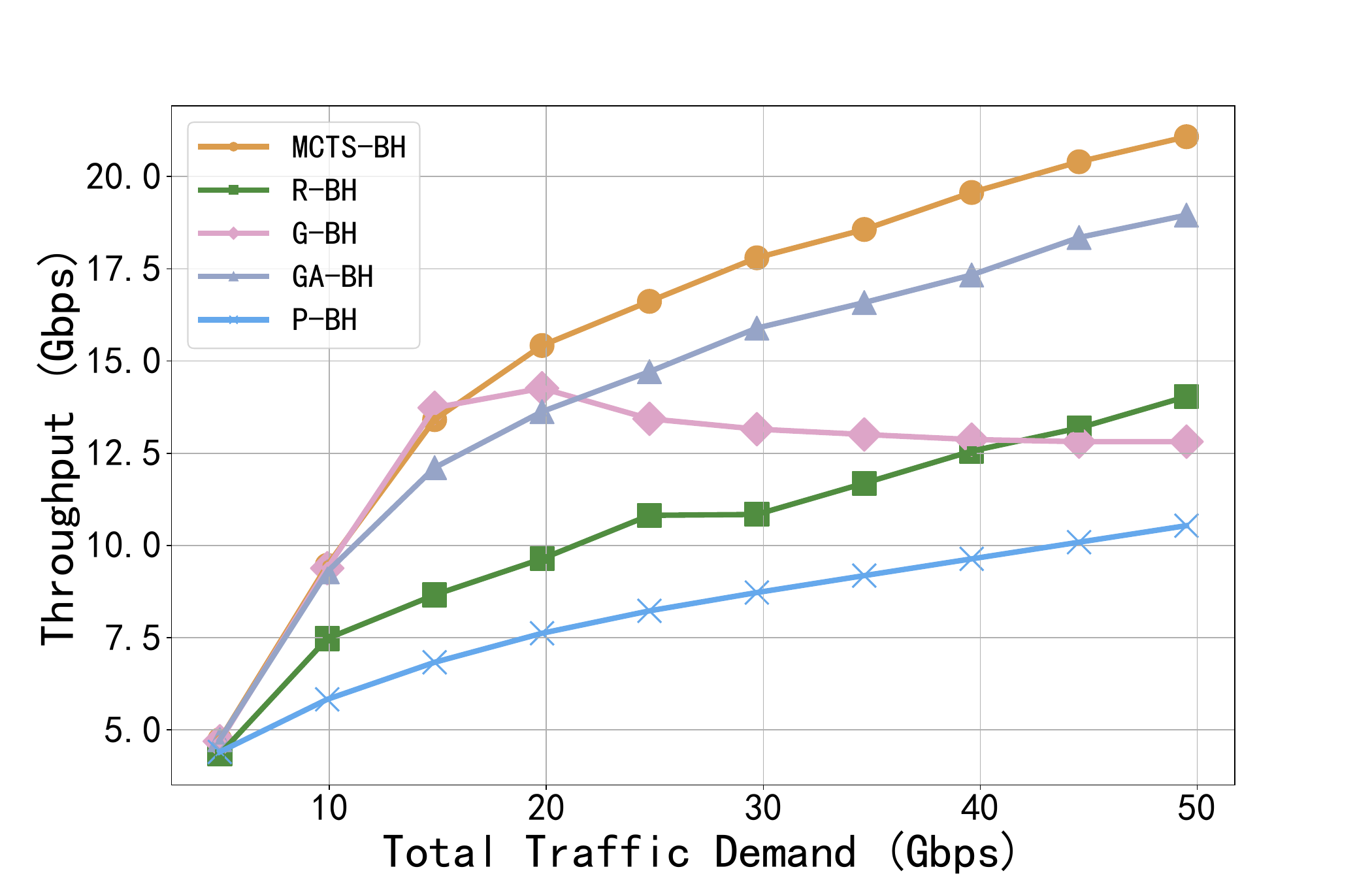}}
    \hfill
    \subfloat[91 cells]{\includegraphics[width=0.31\linewidth]{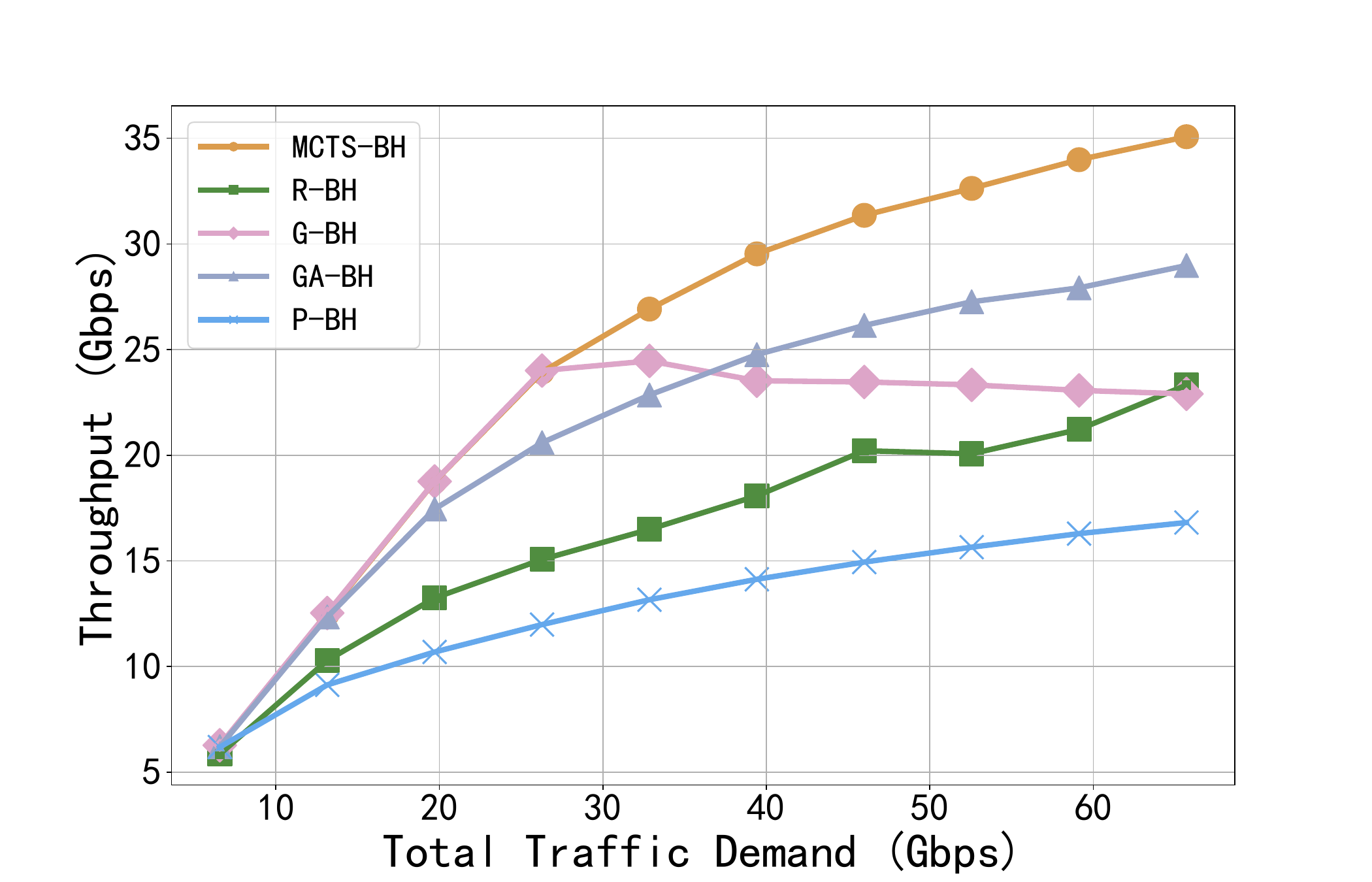}}
    \hfill
    \subfloat[127 cells]{\includegraphics[width=0.31\linewidth]{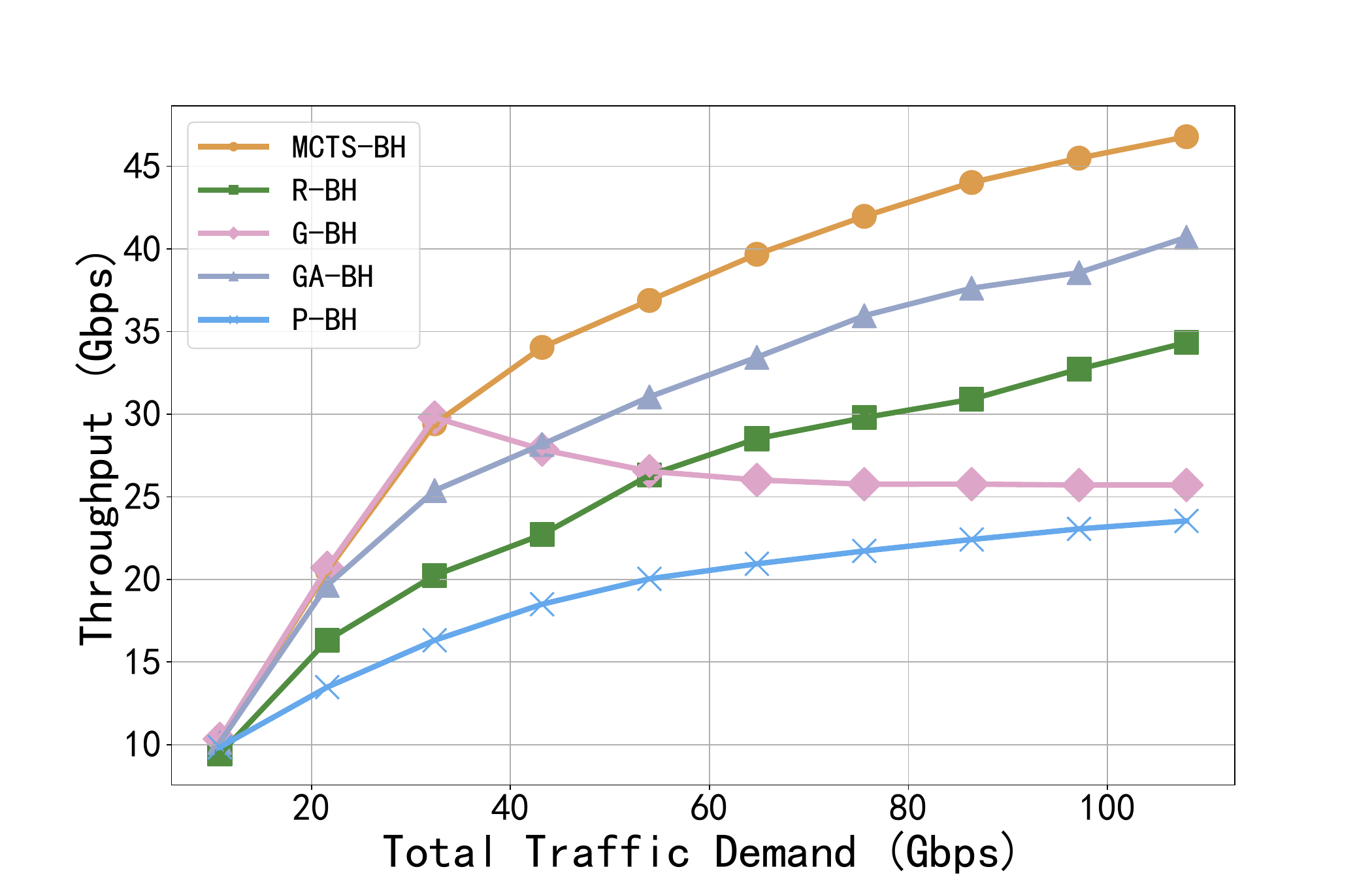}}
    \caption{Throughput performance of different algorithms with different numbers of cells. MCTS-BH (the yellow line) always achieves the highest throughput.}
    \label{fig:big_scale_throughput_performance}
\end{figure*}

We conducted experiments in scenarios with cell numbers and beam numbers of (61, 15), (91, 22), and (127, 31), respectively. In MCTS-BH, the maximum iterations for a single MCTS is set to 200, 300, and 400 to ensure convergence. The experimental results are shown in Fig.~\ref{fig:big_scale_throughput_performance}. In 127 cells, MCTS-BH can increase the throughput by up to 20.85\%, 49.90\% ,81.97\%, and 98.76\% compared with GA-BH, R-BH, G-BH, and P-BH, respectively. 
We can observe that when the traffic demand is low, the capacity of each beam is much greater than the traffic demand, and all algorithms can meet the traffic demand. So the throughput performance of different algorithms is basically consistent. As the demand for traffic increases, simpler methods such as P-BH and R-BH perform poorly because they do not consider traffic demand or CCI. Due to not considering CCI, G-BH exhibits a decrease in performance. GA has good performance, but requires a long computation time. MCTS-BH also exhibits the best performance in terms of high traffic demand. In addition, as the number of cells increases, GA requires more iterations to achieve better performance. Therefore, with a fixed population and generation, the performance of GA worsens with an increasing number of cells.

\subsection{Performance of Computation Time}

Table \ref{tab:BHP computation time} summarizes the time required to compute an illumination pattern in the experiment which is shown in Fig.~\ref{fig:big_scale_throughput_performance}. Among them, P-BH, R-BH, and G-BH can compute illumination patterns within the millisecond timescale, primarily because they adopt relatively simple calculation strategies such as random selection, round-robin scheduling, and greedy algorithms. However, this simplicity also leads to lower overall system throughput. Although MCTS-BH has a computation time in seconds, compared with GA-BH, the average computation time decreases by 81.09\% while achieving a higher throughput performance than GA-BH. This indicates that MCTS-BH is a more suitable algorithm for offline computation than GA-BH.
Based on the computational time analysis, when Tyche employs an online algorithm, G-BH can generate an illumination pattern at the millisecond level. When the corresponding BHTP exists in Tyche's database, the result can be directly retrieved through a database query (approximately 1ms). In either case, Tyche can deliver results within milliseconds, fully demonstrating its advantage in real-time computation.

\begin{table}[t]
    \caption{Computation Time}
    \centering
        \begin{tabular}{|c|c|c|c|c|c|}
        \hline
        \rule[-6pt]{0mm}{16pt} \textbf{Algorithm} & 
        \multicolumn{4}{c|}{\textbf{Computation Time}} \\

        \cline{2-5}
        \rule[-6pt]{0mm}{16pt}
        {} & \textbf{\textit{37 cells}}& \textbf{\textit{61 cells}}& \textbf{\textit{91 cells}} & \textbf{\textit{127 cells}}\\

        \cline{1-5}
        \rule[-6pt]{0mm}{16pt}
        P-BH & 0.003 ms & 0.004 ms & 0.005 ms  & 0.005 ms\\
        \cline{1-5}
        \rule[-6pt]{0mm}{16pt}
        R-BH & 0.016 ms & 0.028 ms & 0.038 ms & 0.049 ms\\
        \cline{1-5}
        \rule[-6pt]{0mm}{16pt}
        G-BH & 0.063 ms & 0.103 ms & 0.272 ms & 0.281 ms \\
        \cline{1-5}
        \rule[-6pt]{0mm}{16pt}
        MCTS-BH  & 12.661 s & 88.070 s & 296.600 s & 905.615 s \\
        \cline{1-5}
        \rule[-6pt]{0mm}{16pt}
        GA-BH  & 330.937 s & 850.870 s & 1836.714 s & 3873.132 s \\
        \cline{1-5}
    \end{tabular}
    \label{tab:BHP computation time}
\end{table}

\subsection{The Impact of MCTS-BH Optimization Algorithm}

As mentioned earlier, we use a scoring algorithm based on the sliding window algorithm and pruning algorithm to optimize the search process of MCTS-BH. Below, we analyze the effectiveness of these two optimization algorithms and the difference in MCTS-BH computation time before and after optimization.
\subsubsection{The Impact of Scoring Algorithm Based on the Sliding Window Algorithm}
\begin{figure}
  \centering
  \includegraphics[width=0.8\linewidth]{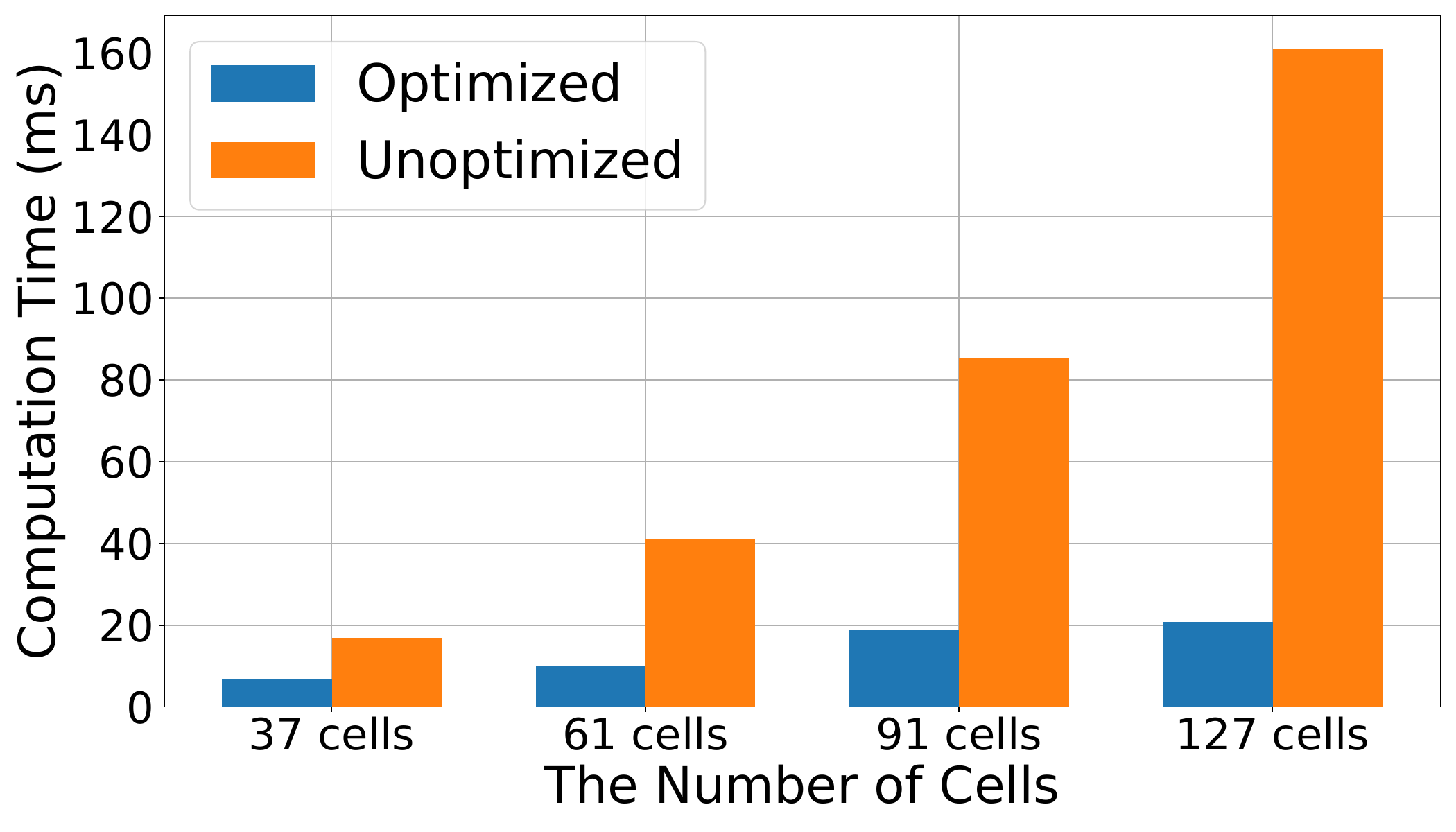}
  \caption{Comparison of the time required for a single score before and after optimization. As the number of cells increases, the optimization effect becomes increasingly obvious.}
  \label{fig:score_time_comparation}
 
\end{figure}

Fig.~\ref{fig:score_time_comparation} shows the time required for a score before and after optimization. After applying the new scoring algorithm, the computation time was reduced by 72.18\% for 127 cells. In the 37 cells, the time difference for one score is not significantly different, and the computation time is reduced by 34.21\%. This is because the interference distance threshold  \(D_{s}\) in the experiment is 1 cells diameter, so almost all service cells in 37 cells are interference cells. So we still need to compute the interference from all other service cells. As the number of the cells increases, the optimization effect becomes increasingly obvious, which also indicates that our optimization algorithm is suitable for large-scale cell scenarios.

\subsubsection{The Impact of Pruning Algorithm}
\begin{figure*}
    \centering
    \subfloat[61 cells \label{fig:figure111label}]{\includegraphics[width=0.31\linewidth]{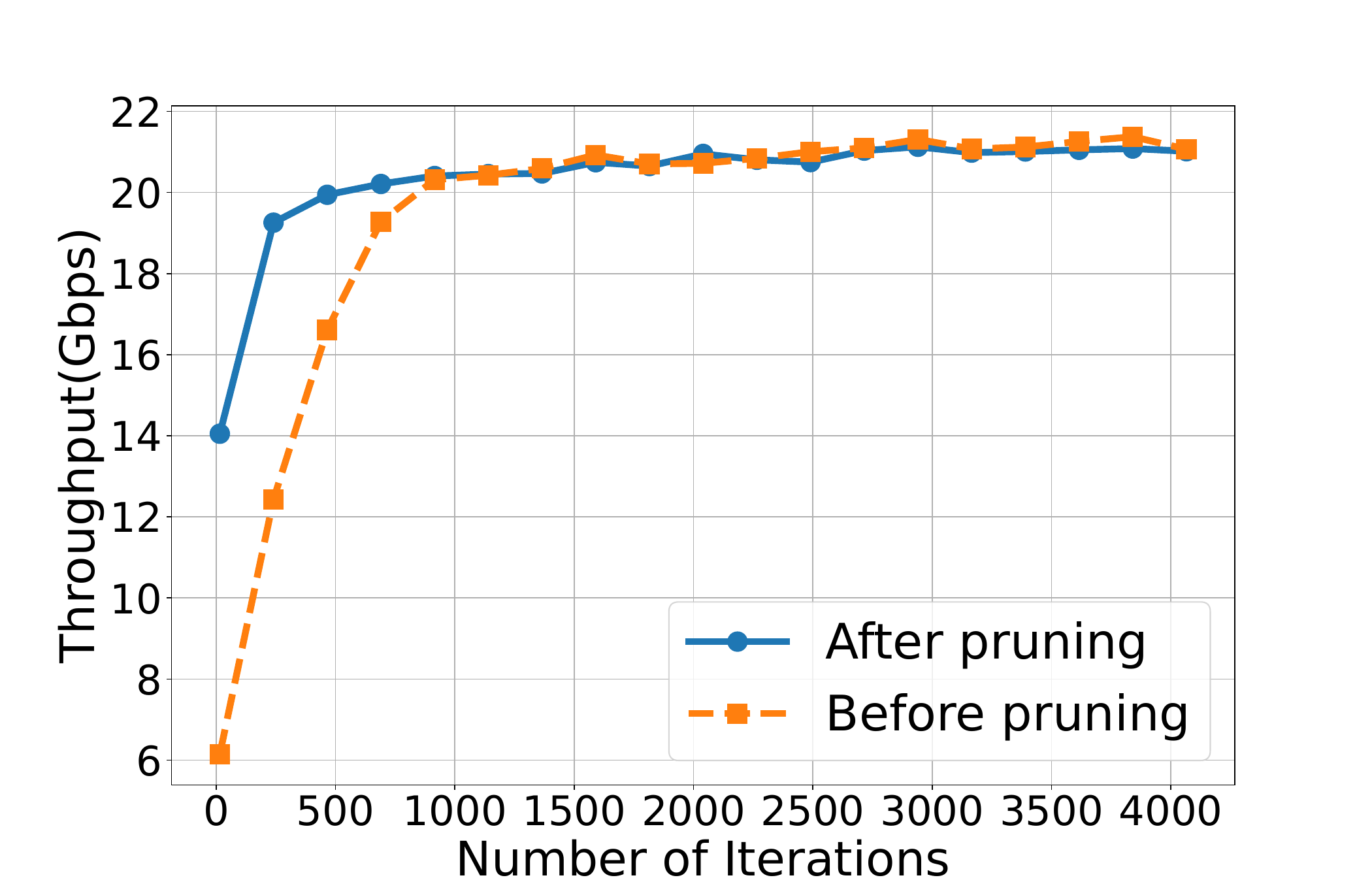}}
    \hfill
    \subfloat[91 cells \label{fig:figure112label}]{\includegraphics[width=0.31\linewidth]{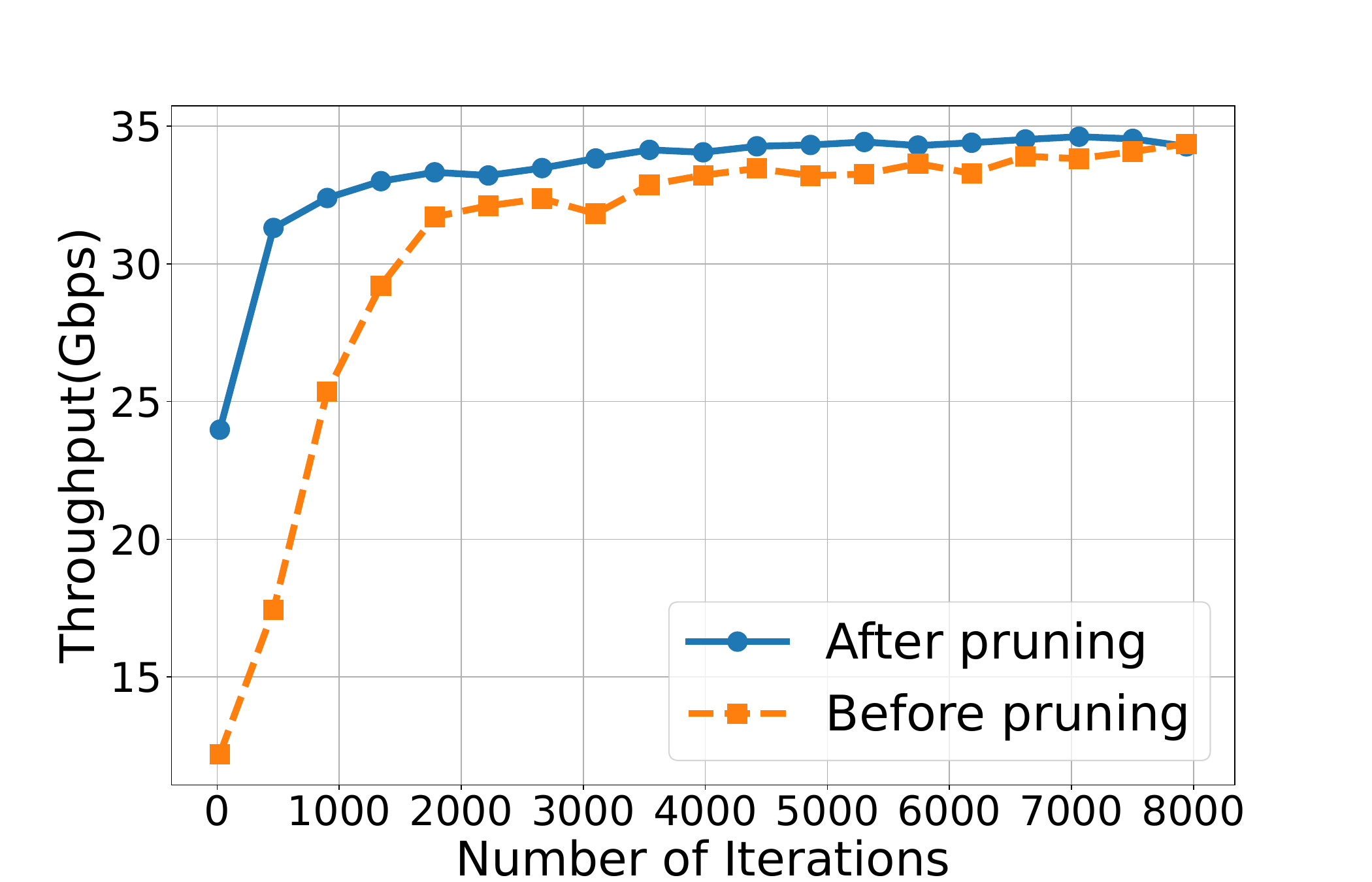}}
    \hfill
    \subfloat[127 cells \label{fig:figure113label}]{\includegraphics[width=0.31\linewidth]{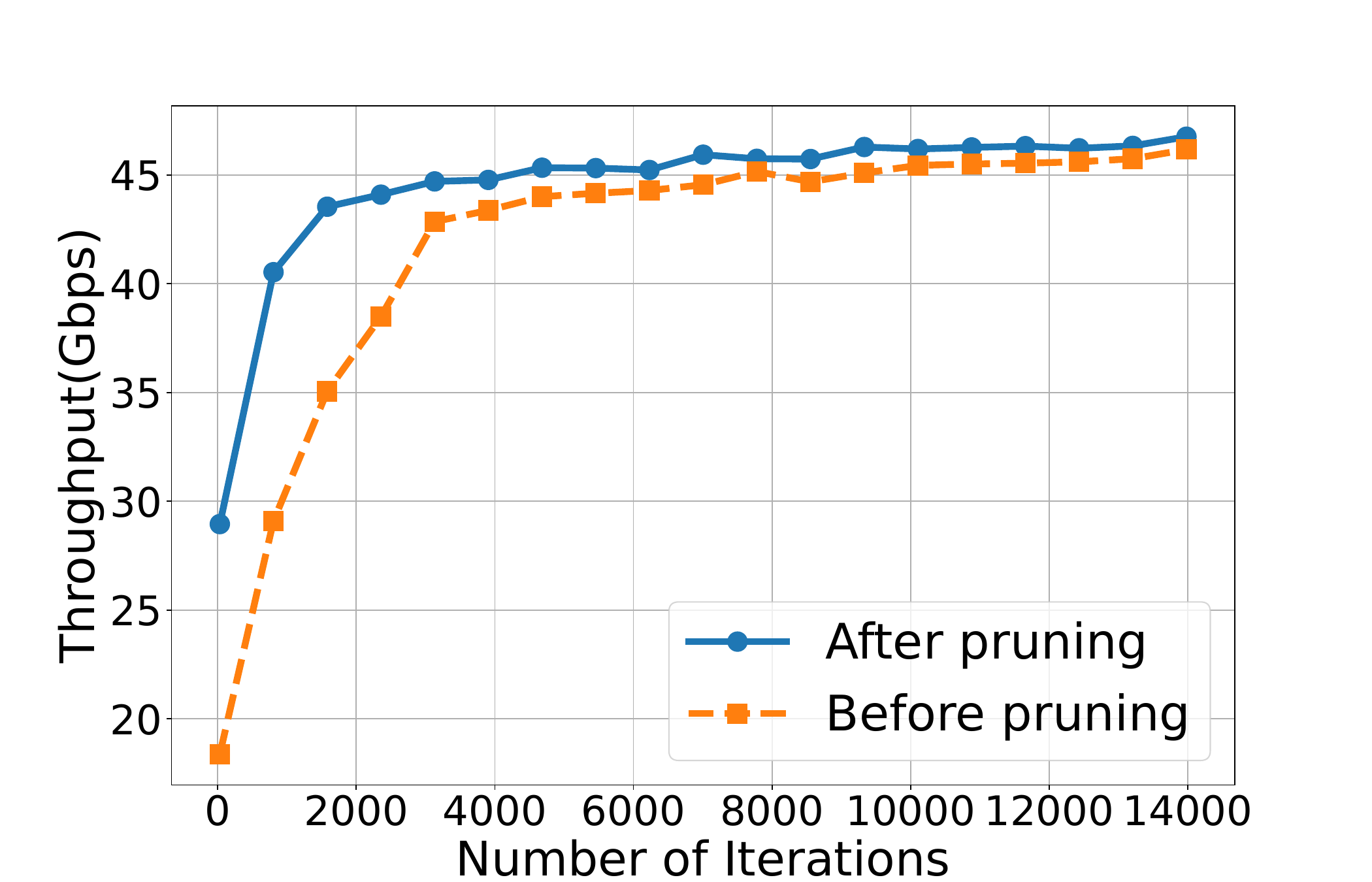}}
    \caption{The change in convergence speed of MCTS-BH after applying the pruning algorithm. After pruning, MCTS-BH can achieve convergence with fewer iterations while maintaining the same throughput performance.}
    \label{fig:Iter_converge}
\end{figure*}

Fig.~\ref{fig:Iter_converge} shows the change in convergence speed of MCTS-BH after applying the pruning algorithm. After applying the pruning algorithm, MCTS-BH can achieve convergence with fewer iterations. For cell counts of 61, 91, and 127, the number of iterations required to achieve convergence decrease by 73.77\%, 74.07\%, and 49.50\%, respectively, while maintaining the same throughput performance.

\subsubsection{Computation Time Comparison}
To illustrate the optimization effects of the two algorithms, we use both the optimized and unoptimized versions of MCTS-BH to compute illumination patterns.
In scenarios with different numbers of cells, we measure the difference in time required to compute an illumination pattern while maintaining approximately the same throughput performance.Table~\ref{tab:Optimized computation time} shows the computation time required to compute an illumination pattern. In scenarios with 37, 61, 91, and 127 cells, the computation time for the optimized MCTS-BH decreased by 41.37\%, 56.94\%, 75.59\%, and 81.41\%, respectively. This demonstrates that our optimization algorithm exhibits significant improvements in large-scale cell scenarios.

 

\begin{figure}
  \centering
  \includegraphics[width=0.8\linewidth]{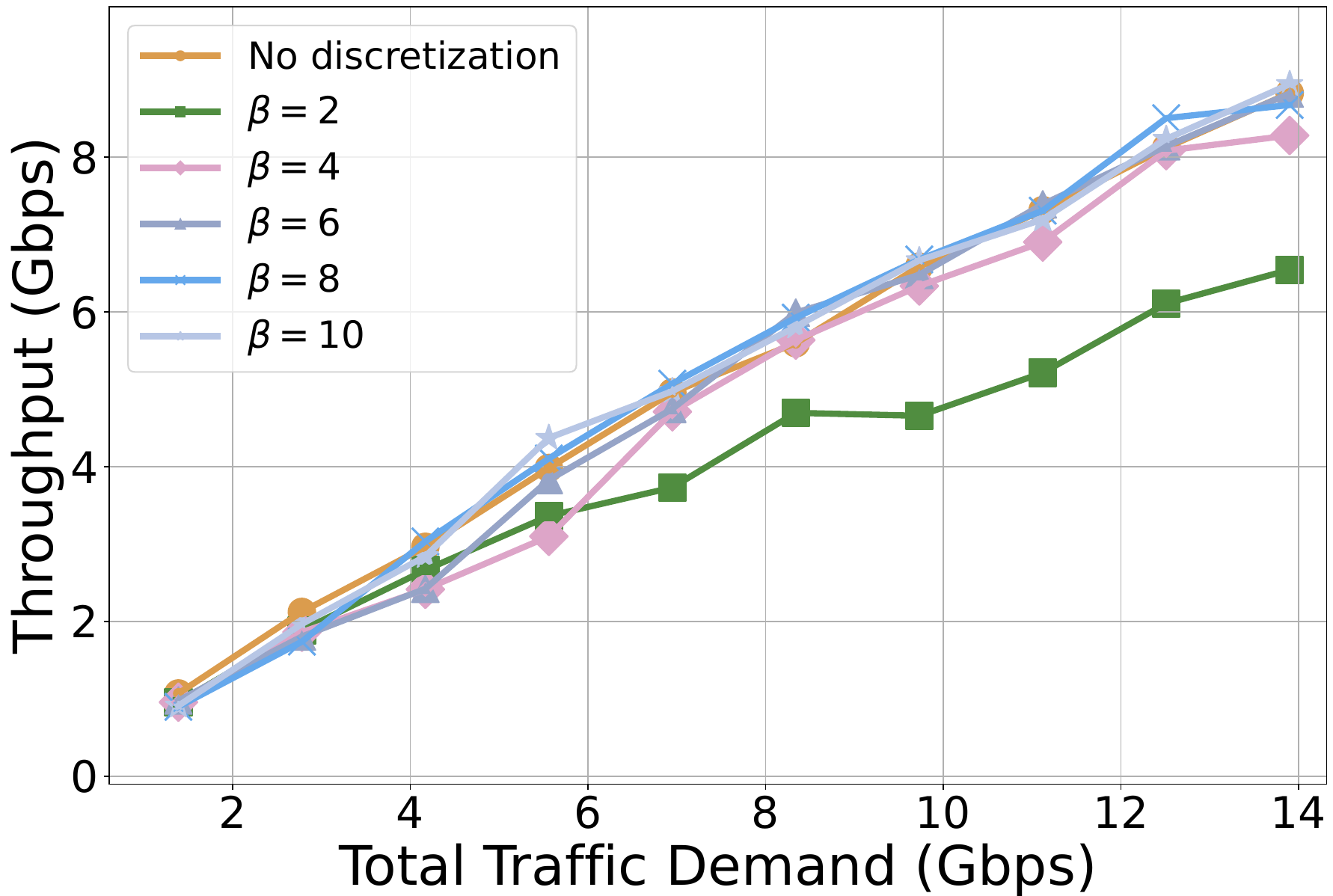}
  \caption{Comparison of MCTS-BH throughput under different discretization factors $\beta$}
  \label{fig:different discretization compare}
\end{figure}

\begin{table}
    \caption{Computation Time of MCTS-BH}
    \centering
        \begin{tabular}{|c|c|c|c|c|c|}
        \hline
        \rule[-6pt]{0mm}{16pt} \textbf{Algorithm} & 
        \multicolumn{4}{c|}{\textbf{Computation Time}} \\

        \cline{2-5}
        \rule[-6pt]{0mm}{16pt}
        {} & \textbf{\textit{37 cells}}& \textbf{\textit{61 cells}}& \textbf{\textit{91 cells}} & \textbf{\textit{127 cells}}\\

        \cline{1-5}
        \rule[-6pt]{0mm}{16pt}
        Unoptimized & 6.723 s & 31.737 s & 158.192 s  & 854.980 s\\
        \cline{1-5}
        \rule[-6pt]{0mm}{16pt}
        Optimized & 3.942 s & 13.665 s & 38.611 s & 158.949 s\\
        \cline{1-5}
        \rule[-6pt]{0mm}{16pt}
        \textbf{\textit{Time Reduction}} & \textbf{41.37 \%} & \textbf{56.94 \%} & \textbf{75.59 \%} & \textbf{81.41 \%}\\
        \cline{1-5}

    \end{tabular}
    \label{tab:Optimized computation time}
\end{table}

\subsection{Impact of Discretization}

This section analyzes the impact of the degree of discretization in the traffic analysis module on the performance of the MCTS-BH algorithm through experiments. The experiments were conducted on 37 cells and 9 beams, with discretization factors $\beta$ set to 2, 4, 6, 8, and 10 respectively. The experimental results are shown in the Fig. \ref{fig:different discretization compare}. The results indicate that when the discretization factor $\beta = 2$, the discretization is too coarse for MCTS-BH to accurately determine the illumination pattern, resulting in a decrease in throughput. However, when $\beta = 4$, the throughput only decreases by 8.4\% compared to the non-decentralized case, showing relatively good performance trade-off. As the discretization factor continues to increase to 6 and above, the throughput is almost identical to that of the non-decentralized situation. These results demonstrate that the proposed discretization strategy can significantly reduce memory usage while only incurring limited throughput loss, thus validating the effectiveness and practicality of the strategy.

\subsection{Memory Consumption Analysis}
This section analyzes the memory usage of the database in the traffic analysis module. To achieve efficient management and retrieval of large-scale data while meeting the system's real-time requirements, we adopt Redis as the database. To evaluate memory usage, we conducted the following analysis.

The structure of a single entry in the database is illustrated in the Fig. \ref{fig:data_structure}. The key, generated by the SHA-256 hash function, occupies 4 bytes. The value consists of the BHTP and the original traffic distribution. The BHTP comprises \( T \) illumination patterns, each containing \( K \) integers of type int, resulting in a memory usage of \( T \times K \times 4 \) bytes for the BHTP. The original traffic distribution includes \( N \) variables of type float, occupying \( 4N \) bytes. Additionally, Redis incurs an overhead of approximately 8 bytes per entry. Thus, the total memory usage of a single entry is \( 4 + T\times K \times 4 + 4N + 8 \) bytes. In the experimental setup of this paper, with \( K = N/4 \), \( T = 30 \), and \( N = 127 \), the memory usage of a single entry is approximately \( 4 + 30 \times (127/4) \times 4 + 4 \times 127 + 8 = 4330 \) bytes. 
Through the discretization operation, taking the discretization factor \( \beta = 4 \) as an example, the theoretical number of traffic distribution states for 127 cells, each with 4 possible states, is \( 4^{127} \approx 1.34 \times 10^{76} \). However, due to constraints imposed by population distribution and usage time, the actual number of traffic patterns occurring in the short term is significantly lower than the theoretical value. Here, we consider storing 200,000 traffic distributions as an example, which requires a total memory of approximately 8.6 GB. This memory demand is entirely manageable for modern high-performance servers, which typically come equipped with 128 GB or more memory, demonstrating the excellent feasibility of the storage structure proposed in this paper in terms of memory consumption. In practical applications, the database size can be flexibly adjusted based on the problem scale and hardware constraints.

\subsection{Impact of Interference Distance Threshold}
\begin{figure}
  \centering
  \includegraphics[width=0.8\linewidth]{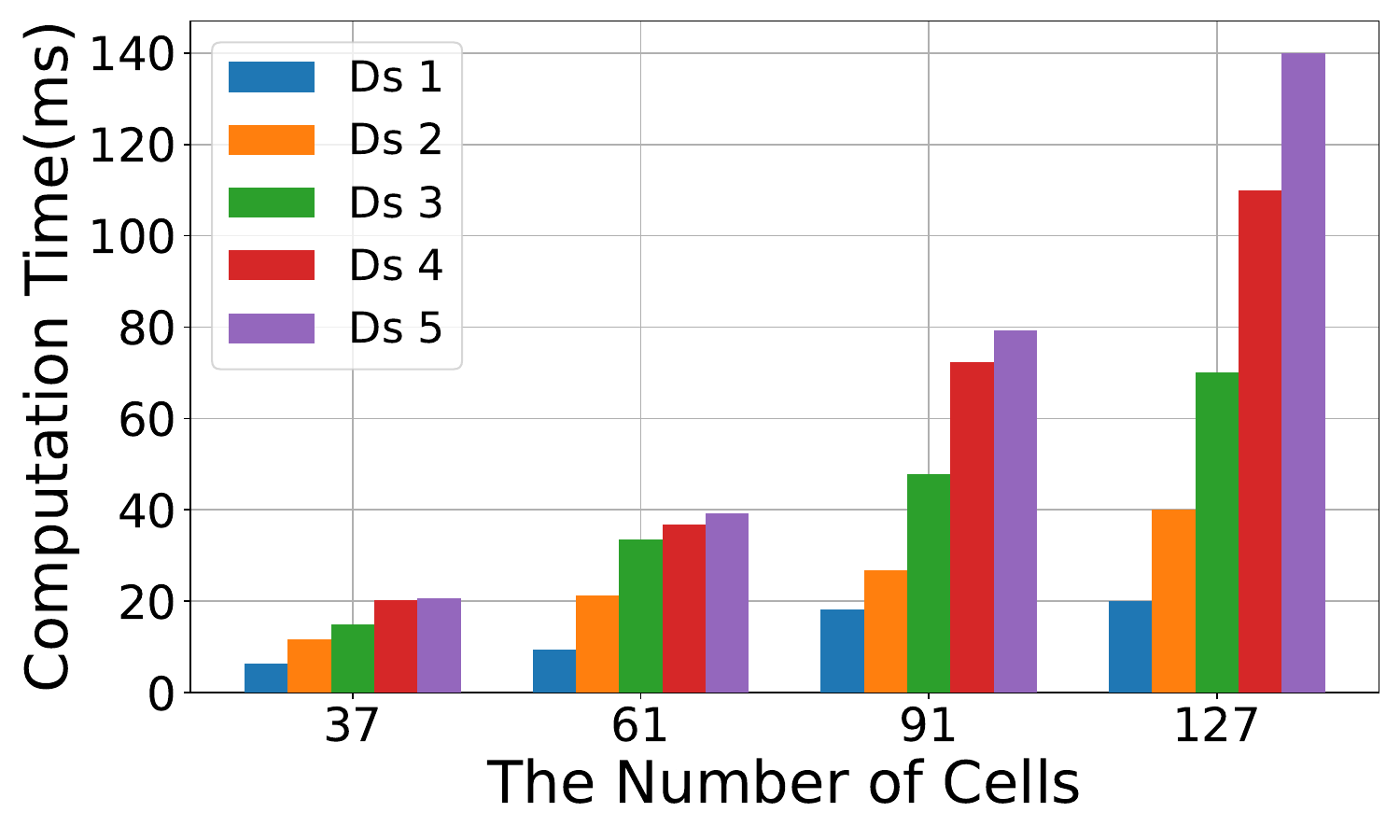}
  \caption{Comparison of MCTS-BH computation time for a single scoring under different \( D_s \) values}
  \label{fig:Ds time compare}
\end{figure}

\begin{figure}
  \centering
  \includegraphics[width=0.8\linewidth]{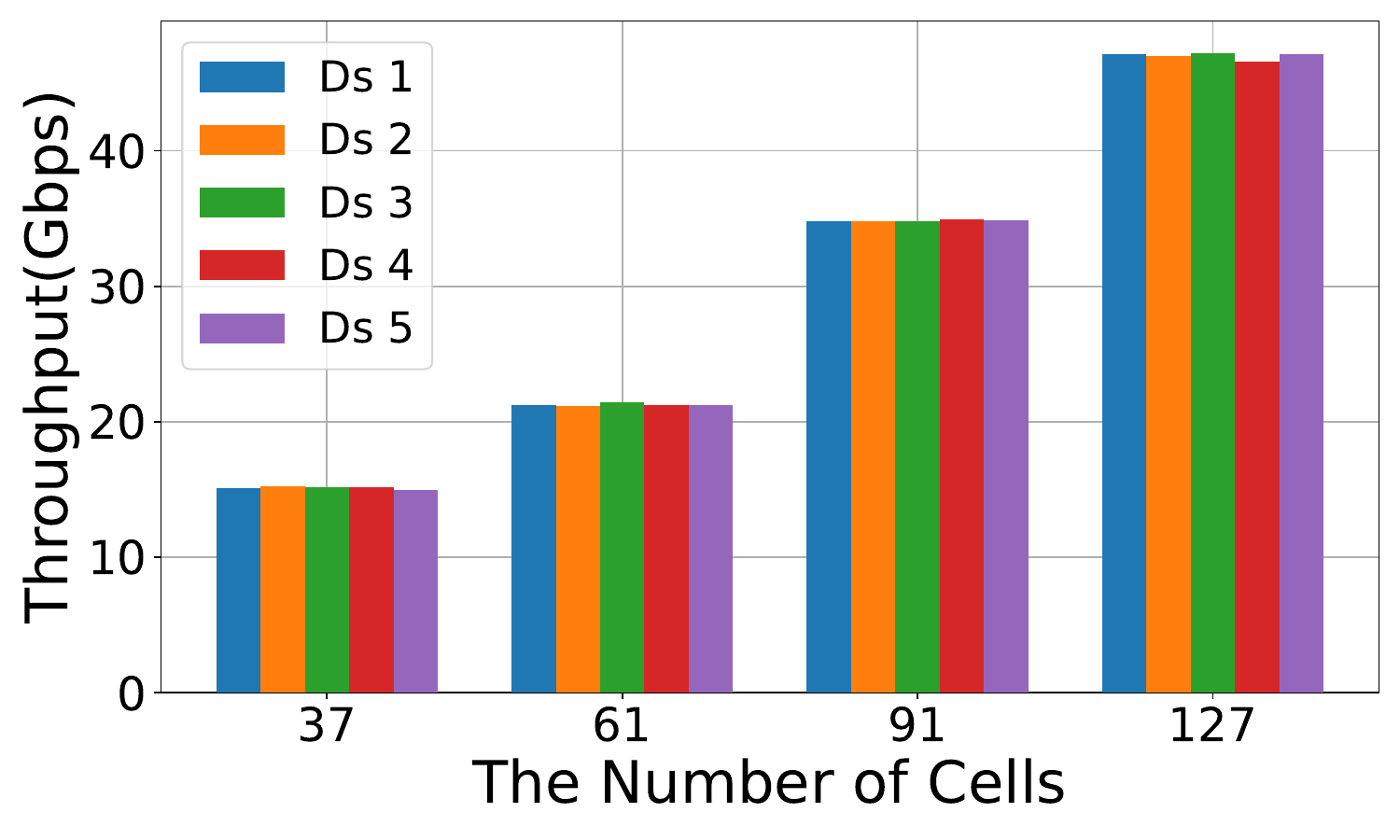}
  \caption{Throughput performance of MCTS-BH under different \( D_s \) values}
  \label{fig:Ds throughput compare}
\end{figure}

This section analyzes the impact of the interference distance threshold \( D_s \) on the algorithm results. \( D_s \) is the interference distance threshold used in the MCTS-BH scoring algorithm. Experiments are conducted under different cell sizes and \( D_s \) values in this section. Figure \ref{fig:Ds time compare} shows the time required for a single scoring computation when \( D_s \) is set to 1, 2, 3, 4, and 5 times the cell diameter. The experimental results indicate that as \( D_s \) increases, the computational time of the algorithm also increases because more interfering cells need to be considered. This effect is most significant when there are 127 cells. Figure \ref{fig:Ds throughput compare} presents the throughput performance of MCTS-BH under different \( D_s \) values. It can be observed that the throughput performance remains similar across different \( D_s \) values due to the low sidelobes of the antenna model used in this study. When using an antenna model with higher sidelobes, increasing \( D_s \) would be necessary to provide more accurate scoring and achieve higher throughput performance.

\section{Future Work}
Currently, Tyche is designed for a single GEO scenario, and we plan to extend it to multi-satellite deployment scenarios in future research. In multi-satellite environments, beam allocation for each satellite must be coordinated based on the global satellite states, introducing two critical challenges: the design of an efficient inter-satellite state synchronization mechanism and the development of an illumination pattern computation algorithm to mitigate inter-satellite interference. The inter-satellite state synchronization mechanism can be achieved either through centralized decision-making by ground stations or by utilizing Inter-Satellite Links (ISL) for direct communication between satellites, coupled with the design of distributed synchronization algorithms to ensure state consistency. To avoid inter-satellite interference, we propose extending the existing Tyche by incorporating the impact of inter-satellite interference into the computation model, such as treating it as a penalty function to optimize the illumination patterns. In summary, designing an efficient inter-satellite synchronization mechanism and an illumination pattern computation algorithm that accounts for inter-satellite interference are core issues to be addressed in multi-satellite scenarios. These directions will be the primary focus of our future research, aiming to enhance satellite system throughput in multi-satellite environments.

\section{CONCLUSION}
In this study, we propose Tyche, a new framework of computing illumination patterns. Specifically, Tyche integrated the following features into the overall design: (1) Tyche uses G-BH for online computation, which can compute an illumination pattern in milliseconds. (2) Tyche uses MCTS-BH for offline computation. Compared with other algorithms, the throughput of MCTS-BH can increase by up to 98.76\%. (3) We use the sliding window algorithm and the pruning algorithm to optimize the search process of MCTS-BH. Compared with the unoptimized algorithm, the computation time for one illumination pattern is reduced by 81.41\% under 127 cells. Finally, we conduct experiments with different numbers of cells to verify the performance advantages of the two algorithms in Tyche, proving that Tyche can be used for real-time illumination pattern computation in large-scale cell scenarios. 

\section*{ACKNOWLEDGMENT}
The research is sponsored by Natural Science Foundation of Shanghai, (Project No. 25ZR1402021)


 
%

\bibliographystyle{IEEEtran}
\bibliography{new_ref}

@inproceedings{BHADV2,
  title={Beam Hopping - a Flexible Satellite Communication System for Mobility},
  author={ Panthi, Sunil  and  Breynaert, Dirk  and  Mclain, Christopher  and  King, Janet },
  booktitle={Aiaa International Communications Satellite Systems Conference},
  year={2017},
}

@ARTICLE{NPref,
  author={Chen, Lin and Ha, Vu Nguyen and Lagunas, Eva and Wu, Linlong and Chatzinotas, Symeon and Ottersten, Björn},
  journal={IEEE Transactions on Wireless Communications}, 
  title={The Next Generation of Beam Hopping Satellite Systems: Dynamic Beam Illumination With Selective Precoding}, 
  year={2023},
  volume={22},
  number={4},
  pages={2666-2682},
  doi={10.1109/TWC.2022.3213418}}

@ARTICLE{BH_INCREASE_Th2,
  author={Li, Yitao and Luo, Zhongqiang and Zhou, Wuyang and Zhu, Jinkang},
  journal={China Communications}, 
  title={Benefits analysis of beam hopping in satellite mobile system with unevenly distributed traffic}, 
  year={2021},
  volume={18},
  number={9},
  pages={11-23},
  doi={10.23919/JCC.2021.09.002}}

@article{ZhanglEO,
  title={System-Level Evaluation of Beam Hopping in NR-Based LEO Satellite Communication System},
  author={Jingwei Zhang and Dali Qin and Chuili Kong and Feiran Zhao and Rong Li and Jun Wang},
  journal={2023 IEEE Wireless Communications and Networking Conference (WCNC)},
  year={2022},
  pages={1-6},
  url={https://api.semanticscholar.org/CorpusID:248987727}
}

@inproceedings{SNS3,
  title={System Level Modeling of Beam Hopping for Multi-Spot Beam Satellite Systems},
  author={ Sormunen, Lauri  and  Puttonen, Jani  and  Kurjenniemi, Janne },
  booktitle={23rd Ka and Broadband Communications Conference (KaConf)},
  year={2017},
}

@inproceedings{GA1,
  title={Beam Hopping in Multi-Beam Broadband Satellite Systems: System Performance and Payload Architecture Analysis},
  author={ Angeletti, Piero  and  Prim, Fernandez D  and  Rinaldo, Rita },
  booktitle={Aiaa, San Diego},
  year={2006},
}

@ARTICLE{DRL1,
  author={Hu, Xin and Zhang, Yuchen and Liao, Xianglai and Liu, Zhijun and Wang, Weidong and Ghannouchi, Fadhel M.},
  journal={IEEE Transactions on Broadcasting}, 
  title={Dynamic Beam Hopping Method Based on Multi-Objective Deep Reinforcement Learning for Next Generation Satellite Broadband Systems}, 
  year={2020},
  volume={66},
  number={3},
  pages={630-646},
  doi={10.1109/TBC.2019.2960940}
}

@ARTICLE{DRL2,
  author={Hu, Xin and Wang, Libing and Wang, Yin and Xu, Sujie and Liu, Zhijun and Wang, Weidong},
  journal={IEEE Communications Letters}, 
  title={Dynamic Beam Hopping for DVB-S2X GEO Satellite: A DRL-Powered GA Approach}, 
  year={2022},
  volume={26},
  number={4},
  pages={808-812},
  doi={10.1109/LCOMM.2022.3141420}}

@INPROCEEDINGS{DRL3,
  author={Wang, Haonan and Liu, Lixiang and Zhou, Xin and Xu, Lexi and Wu, Guangyang and Liu, Shuaijun},
  booktitle={2023 IEEE 22nd International Conference on Trust, Security and Privacy in Computing and Communications (TrustCom)}, 
  title={Deep Reinforcement Learning Based Interference Avoidance Beam-Hopping Allocation Algorithm in Multi-beam Satellite Systems}, 
  year={2023},
  volume={},
  number={},
  pages={1966-1973},
  doi={10.1109/TrustCom60117.2023.00268}}

@ARTICLE{MADRL,
  author={Lin, Zhiyuan and Ni, Zuyao and Kuang, Linling and Jiang, Chunxiao and Huang, Zhen},
  journal={IEEE Transactions on Vehicular Technology}, 
  title={Dynamic Beam Pattern and Bandwidth Allocation Based on Multi-Agent Deep Reinforcement Learning for Beam Hopping Satellite Systems}, 
  year={2022},
  volume={71},
  number={4},
  pages={3917-3930},
  doi={10.1109/TVT.2022.3145848}}

@article{alphaGo,
  title={Mastering the game of Go with deep neural networks and tree search},
  author={David Silver and Aja Huang and Chris J. Maddison and Arthur Guez and L. Sifre and George van den Driessche and Julian Schrittwieser and Ioannis Antonoglou and Vedavyas Panneershelvam and Marc Lanctot and Sander Dieleman and Dominik Grewe and John Nham and Nal Kalchbrenner and Ilya Sutskever and Timothy P. Lillicrap and Madeleine Leach and Koray Kavukcuoglu and Thore Graepel and Demis Hassabis},
  journal={Nature},
  year={2016},
  volume={529},
  pages={484-489},
}

@ARTICLE{DVB-S2X,
  journal={ETSI Std}, 
  title={Digital Video Broadcasting (DVB); Implementation guidelines for the second generation system for Broadcasting, Interactive Services, News Gathering and other broadband satellite applications; Part 2: S2 Extensions (DVB-S2X)}, 
  year={2021},
}

@article{MINLP,
  title={Mixed-integer nonlinear optimization*†},
  author={Pietro Belotti and Christian Kirches and Sven Leyffer and Jeff T. Linderoth and James R. Luedtke and Ashutosh Mahajan},
  journal={Acta Numerica},
  year={2013},
  volume={22},
  pages={1 - 131},
  url={https://api.semanticscholar.org/CorpusID:122995288}
}

@Misc{JUPITER3,
	howpublished = {[Online]},
	note = {\url{https://www.hughes.com/what-we-offer/satellite-services/jupiter-geo-satellites/JUPITER3}},
	title = {JUPITER3 Web},
}

@Misc{Viasat,
	howpublished = {[Online]},
	note = {\url{https://www.viasat.com/space-innovation/satellite-fleet/viasat-2/}},
	title = {Viasat Web},
}

@Misc{Konnect,
	howpublished = {[Online]},
	note = {\url{https://www.eutelsat.com/en/satellites/2-7-east.html}},
	title = {Konnect Web},
}

@INPROCEEDINGS{DyBhQifa1,
  author={Wu, Zhenguo and Ren, Pinyi and Xu, Dongyang},
  booktitle={2022 IEEE 95th Vehicular Technology Conference: (VTC2022-Spring)}, 
  title={Flexible Resource Allocation for Differentiated QoS Provisioning in Beam-Hopping Satellite Communications System}, 
  year={2022},
  volume={},
  number={},
  pages={1-5},
  doi={10.1109/VTC2022-Spring54318.2022.9860761}}

@Article{DyBhQifa2,
AUTHOR = {Gou, Liang and Bian, Dongming and Dong, Baogui and Nie, Yulei},
TITLE = {Improved Spread Spectrum Aloha Protocol and Beam-Hopping Approach for Return Channel in Satellite Internet of Things},
JOURNAL = {Sensors},
VOLUME = {23},
YEAR = {2023},
NUMBER = {4},
ARTICLE-NUMBER = {2116},
URL = {https://www.mdpi.com/1424-8220/23/4/2116},
PubMedID = {36850716},
ISSN = {1424-8220},
DOI = {10.3390/s23042116}
}

@Article{DyBhGA,
AUTHOR = {Guo, Shengjun and Han, Kai and Gong, Wenbin and Li, Lu and Tian, Feng and Jiang, Xinglong},
TITLE = {An Efficient Multi-Dimensional Resource Allocation Mechanism for Beam-Hopping in LEO Satellite Network},
JOURNAL = {Sensors},
VOLUME = {22},
YEAR = {2022},
NUMBER = {23},
ARTICLE-NUMBER = {9304},
PubMedID = {36502006},
ISSN = {1424-8220},
DOI = {10.3390/s22239304}
}

@Misc{H3,
	howpublished = {[Online]},
	note = {\url{https://h3geo.org/docs/}},
	title = {H3 Information},
	}

@ARTICLE{3gpp,
  journal={3GPP, document TR 38.811}, 
  title={Study on New Radio (NR) to Support Non-Terrestrial Networks}, 
  year={2019},
}

@ARTICLE{MCTS,
  author={Browne, Cameron B. and Powley, Edward and Whitehouse, Daniel and Lucas, Simon M. and Cowling, Peter I. and Rohlfshagen, Philipp and Tavener, Stephen and Perez, Diego and Samothrakis, Spyridon and Colton, Simon},
  journal={IEEE Transactions on Computational Intelligence and AI in Games}, 
  title={A Survey of Monte Carlo Tree Search Methods}, 
  year={2012},
  volume={4},
  number={1},
  pages={1-43},
}

@inproceedings{HTS,
  title={Evolution of High Throughput Satellite Systems: Vision, Requirements, and Key Technologies},
  author={Olfa Ben Yahia and Zineb Garroussi and Olivier B'elanger and Brunilde Sans{\`o} and Jean-François Frigon and St{\'e}phane Martel and Antoine Lesage-Landry and Gunes Karabulut Kurt},
  year={2023},
  url={https://api.semanticscholar.org/CorpusID:263828967}
}

@INPROCEEDINGS{CCI,
  author={Lutz, Erich},
  booktitle={2015 IEEE International Conference on Communications (ICC)}, 
  title={Co-channel interference in high-throughput multibeam satellite systems}, 
  year={2015},
  volume={},
  number={},
  pages={885-891},
  doi={10.1109/ICC.2015.7248434}}

@ARTICLE{DRLGA,
  author={Hu, Xin and Wang, Libing and Wang, Yin and Xu, Sujie and Liu, Zhijun and Wang, Weidong},
  journal={IEEE Communications Letters}, 
  title={Dynamic Beam Hopping for DVB-S2X GEO Satellite: A DRL-Powered GA Approach}, 
  year={2022},
  volume={26},
  number={4},
  pages={808-812},
  doi={10.1109/LCOMM.2022.3141420}}

@ARTICLE{BHADV,
  author={Fourati, Fares and Alouini, Mohamed-Slim},
  journal={Intelligent and Converged Networks}, 
  title={Artificial intelligence for satellite communication: A review}, 
  year={2021},
  volume={2},
  number={3},
  pages={213-243},
  doi={10.23919/ICN.2021.0015}}

@online{RedisBenchmarking,
  author = {{Redis Community}},
  title = {Redis Benchmarking},
  year = {2023},
  url = {https://redis.io/docs/management/optimization/benchmarks/},
  urldate = {2025-04-01},
  organization = {Redis Ltd.}
}

@ARTICLE{Plotinus,
  author={Gao, Yue and Qiu, Kun and Chen, Zhe and Zhu, Wenjun and Zhang, Qi and Luo, Handong and Lin, Quanwei and Yang, Ziheng and Liu, Wenhao},
  journal={Journal of Communications and Information Networks}, 
  title={Plotinus: A Satellite Internet Digital Twin System}, 
  year={2024},
  volume={9},
  number={1},
  pages={24-33},
  doi={10.23919/JCIN.2024.10494942}}

@ARTICLE{antenna,
  author={Lin, Zhiyuan and Ni, Zuyao and Kuang, Linling and Jiang, Chunxiao and Huang, Zhen},
  journal={IEEE Transactions on Communications}, 
  title={Multi-Satellite Beam Hopping Based on Load Balancing and Interference Avoidance for NGSO Satellite Communication Systems}, 
  year={2023},
  volume={71},
  number={1},
  pages={282-295},
  doi={10.1109/TCOMM.2022.3226190}}

@techreport{3GPP2023,
  author = {{3GPP}},
  title = {{New WID: Non-Terrestrial Networks (NTN) for NR Phase 3}},
  institution = {{3rd Generation Partnership Project (3GPP)}},
  number = {{TSG RAN \#102, RP-234078}},
  year = {2023}
}

@INPROCEEDINGS{11148638,
  author={Yang, Chunyu and Yang, Boyu and Qiu, Kun and Chen, Zhe and Gao, Yue},
  booktitle={2025 IEEE/CIC International Conference on Communications in China (ICCC)}, 
  title={DualAttWaveNet: Multiscale Attention Networks for Satellite Interference Detection}, 
  year={2025},
  volume={},
  number={},
  pages={1-6},
  doi={10.1109/ICCC65529.2025.11148638}}



\begin{IEEEbiography}[{\includegraphics[width=1in,height=1.25in,clip,keepaspectratio]{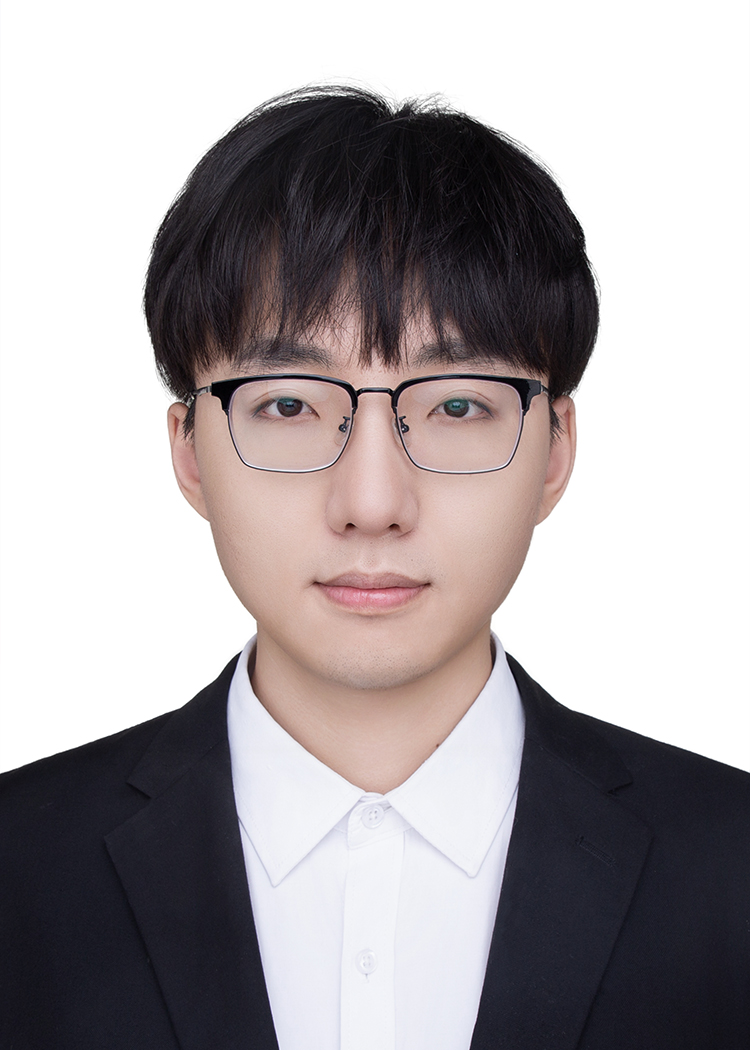}}]{Ziheng Yang}
earned his bachelor’s degree in communication engineering from Nanjing University of Posts and Telecommunications in 2022. He is pursuing his master’s degree at the School of Computer Science, Fudan University, Shanghai, China. His research interests include satellite communication and beam hopping.
\end{IEEEbiography}
\begin{IEEEbiography}[{\includegraphics[width=1in,height=1.25in,clip,keepaspectratio]{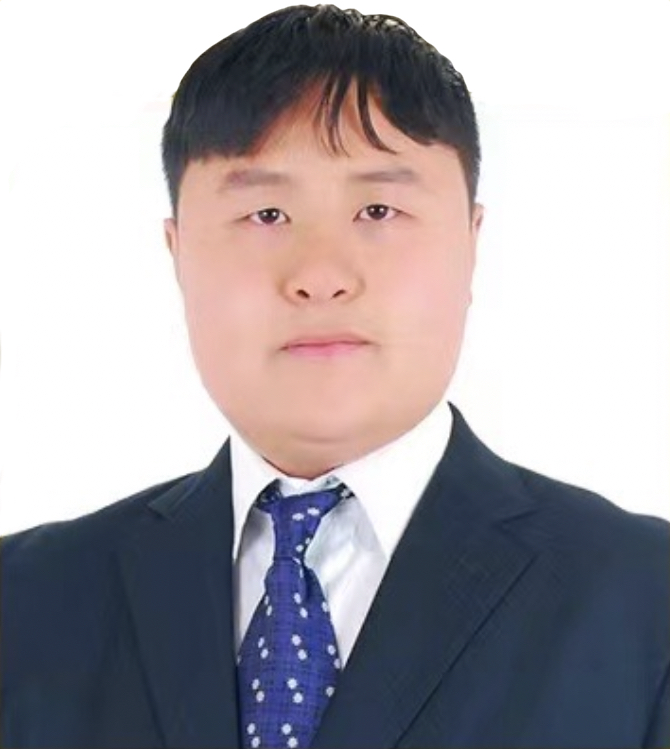}}]{Kun Qiu}(Senior Member, IEEE) [corresponding author] received his B.Sc. from Fudan University in 2013 and his PhD from Fudan University in 2019. He works for Intel as a software engineer from 2019 to 2023. He joined Fudan University in 2023 as an Assistant Professor in the School of Computer Science at Fudan University. His research interests include computer networks and computer architecture. He is a senior member of IEEE, CCF and a member of ACM.
\end{IEEEbiography}
\begin{IEEEbiography}[{\includegraphics[width=1in,height=1.25in,clip,keepaspectratio]{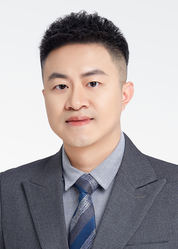}}]{Zhe Chen}
(Member,IEEE) received his PhD in Computer Science from Fudan University, China, with a 2019 ACM SIGCOMM China Doctoral Dissertation Award. He is an Assistant Professor in the School of Computer Science at Fudan University and the Co-Founder of AIWiSe Ltd. Inc. Before joining Fudan University, he worked as a research fellow at NTU for three years, and his research achievements, along with his efforts in launching products based on them, have thus earned him 2021 ACM SIGMOBILE China Rising Star Award recently.
\end{IEEEbiography}
\begin{IEEEbiography}[{\includegraphics[width=1in,height=1.25in,clip,keepaspectratio]{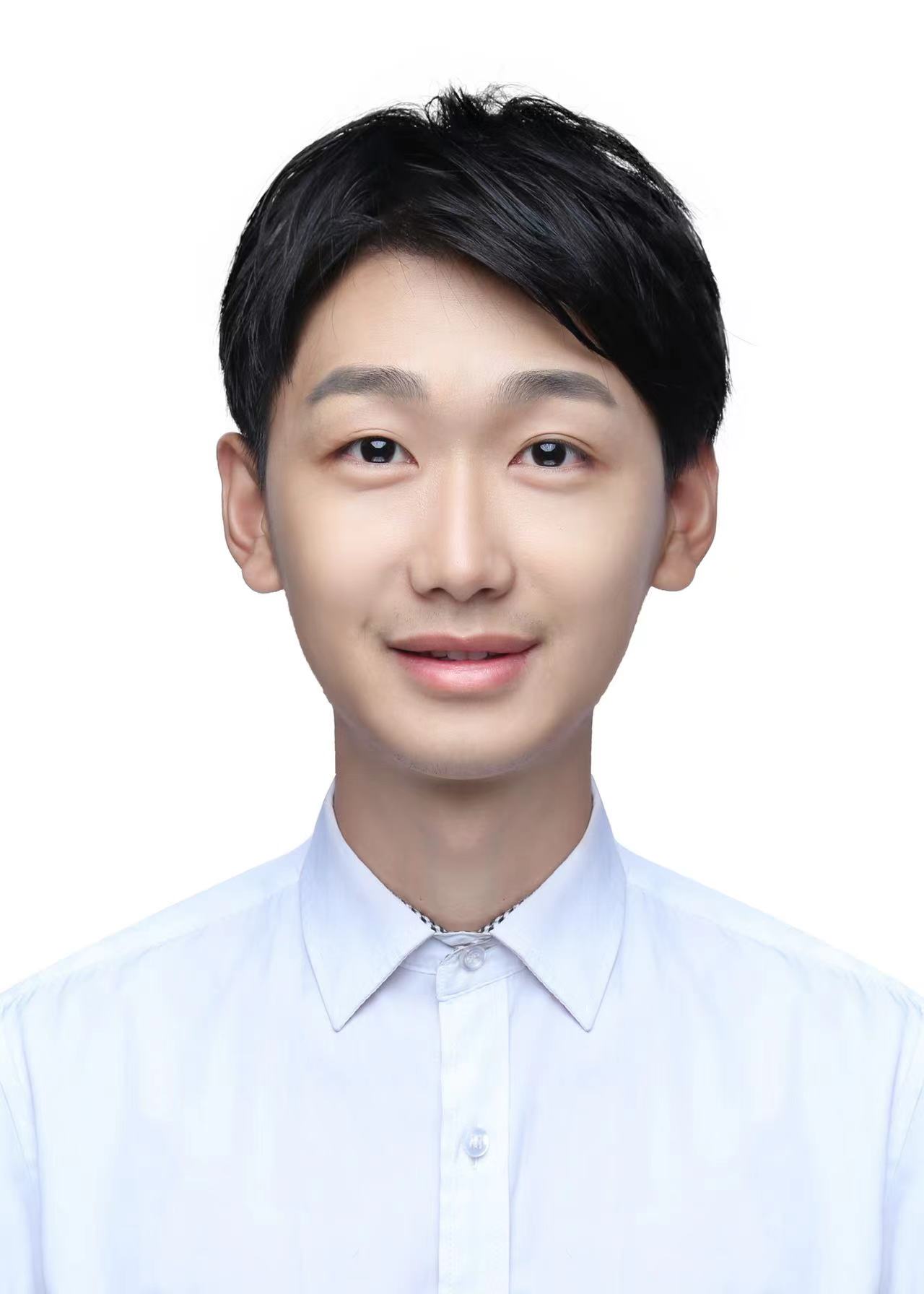}}]{Wenjun Zhu}(Member, IEEE) received his master's Degree from NanJing University of Posts and Telecommunications in 2020. He works for Intel as a software engineer from 2020 to 2023. Now he joined Fudan University in 2023 as a software engineer.His research interests include computer networks,computer architecture and cloud-native. He is also a member of CCF.
\end{IEEEbiography}
\begin{IEEEbiography}[{\includegraphics[width=1in,height=1.25in,clip,keepaspectratio]{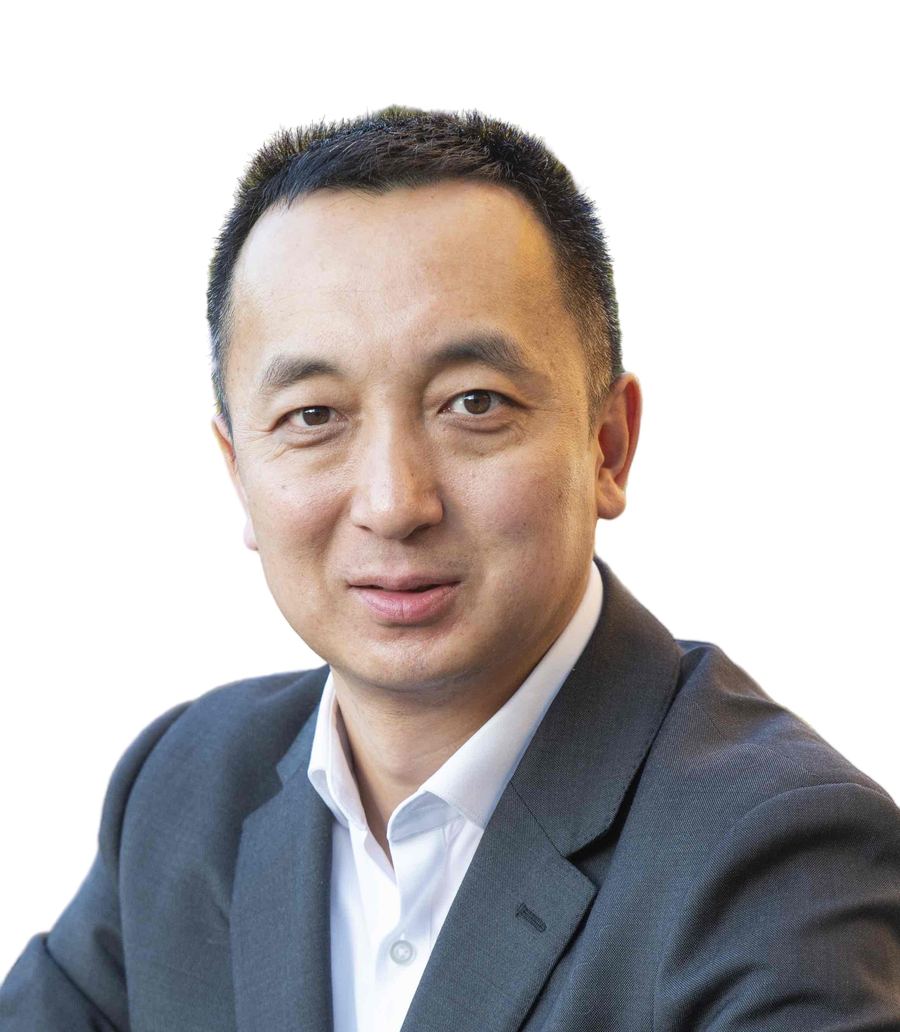}}]{Yue Gao}(Fellow, IEEE) is a Chair Professor and Dean of the Institute of Space Internet at Fudan University, China. He is a Fellow of the IEEE, the IET and CIC. He received his MSc and PhD from Queen Mary University of London (QMUL), UK, in 2003 and 2007. He worked as a faculty staff member, professor, and chair professor at QMUL and the University of Surrey in the U.K. He was a co-recipient of the EU Horizon Prize Award on Collaborative Spectrum Sharing and elected an Engineering and Physical Sciences Research Council Fellow.
\end{IEEEbiography}

\vspace{11pt}

\vfill

\end{document}